\documentclass[preprint,preprintnumbers,nofootinbib,superscriptaddress]{revtex4-1}

\usepackage{amsmath,amsfonts,amssymb}

\usepackage{bm}
\usepackage{bbm}
\usepackage{mathrsfs}

\usepackage{tikz-cd}
\usepackage[all,cmtip]{xy}
\xyoption{rotate}

\begin{document}
\preprint{KEK-TH-1962}
\title{On Quantisations, Quasi-probabilities and the Weak Value}
\author{Jaeha Lee}
\affiliation{National Institute of Informatics, 2-1-2 Hitotsubashi, Chiyoda-ku, Tokyo 101-8430, Japan}
\email{jlee@post.kek.jp}
\author{Izumi Tsutsui}
\affiliation{Theory Center, Institute of Particle and Nuclear Studies, High Energy Accelerator Research Organization (KEK), 1-1 Oho, Tsukuba, Ibaraki 305-0801, Japan}
\email{izumi.tsutsui@kek.jp}
\date{\today}
\begin{abstract}
We propose a general framework of the quantum/quasi-classical transformations by introducing the concept of quasi-joint-spectral distribution (QJSD).  Specifically, we show that the QJSDs uniquely yield various pairs of quantum/quasi-classical transformations, including the Wigner-Weyl transform.
We also discuss the statistical behaviour of combinations of generally non-commutin quantum observables by introducing the concept of quantum correlations and conditional expectations defined analogously to the classical counterpart.  Based on these, Aharonov's weak value is given a statistical interpretation as one realisation of the quantum conditional expectations furnished in our formalism.
\end{abstract}

\maketitle

\section{Introduction}

Since the advent of quantum theory founded nearly a century ago,
non-commutativity of quantum observables has undoubtedly been in the centrepiece of the theory marking its departure from classical theory.  
The hallmark of this is Heisenberg's uncertainty relation \cite{Heisenberg_1927}, which has later been elaborated from operational viewpoints by taking account of the measurement device by Ozawa \cite{Ozawa_2003, Ozawa_2004}.   
At the same time, the non-commutativity has been one of the major sources of troubles we face when we try to interpret their measurement outcomes in a sensible manner.  This has naturally led to various attempts of `quantisation' of classical systems, most notably in terms of non-commuting Hilbert space operators, or conversely of `quasi-classical' interpretation of quantum systems in terms of commuting quantities familiar to us in classical theory.
  
The study on quantum and quasi-classical transformations has a long history dating back to the early days of quantum mechanics.  Wigner and Weyl were among the prominent figures who have made much contribution in this effort bearing the theory of Wigner-Weyl transform \cite{Weyl_1927, Wigner_1932}.  Historically, however, all these contributions in this area have been made more or less in a heuristic manner, and apparently their systematic treatment is still underdeveloped, not to mention a transparent overview of the relations among the various proposals of the transformations made so far.

On the other hand, in recent years we have witnessed the rise of interest in an issue related, at the roots, to the interpretation of measurement outcomes under non-commutativity.  It is the novel quantity called the \emph{weak value},
\begin{equation}
A_{w} := \frac{\langle \psi^{\prime}, A \psi \rangle}{\langle \psi^{\prime}, \psi \rangle}
\end{equation}
which has been proposed by Aharonov and co-workers \cite{Aharonov_1988} based on their time-symmetric formulation of quantum mechanics \cite{Aharonov_1964}.  The weak value
is a physical quantity that characterises the value of the observable $A$ in the process specified by an initial (pre-selected) state $\vert\psi \rangle$ and a final (post-selected) state $\vert\psi^{\prime} \rangle$.  Unlike the standard measurement outcomes given by one of the eigenvalues of an observable $A$ obtained in an ideal measurement, the weak value admits a definite value and is considered to be meaningful even for a set of non-commutable observables.  The relation between the weak value and the quasi-classical transformations has been argued earlier, specifically with the Kirkwood-Dirac distribution \cite{Ozawa_2011, Hofmann_2014}.  

One of the aims of our present paper, expounded in Section~\ref{sec:two}, is to propose a general framework of the quantum/quasi-classical transformations by introducing the concept of \emph{quasi-joint-spectral distribution} (QJSD).  Specifically, we show that the QJSDs, of which definition shortly follows, uniquely yield various pairs of quantum/quasi-classical transformations, and that notable previous proposals of the transformations belong to this framework as special cases.  Another aim, to which Section~\ref{sec:three} is devoted, is to discuss the statistical behaviour of combinations of \emph{generally non-commuting} quantum observables.  Specifically, we introduce the concept of quantum correlations and conditional expectations, which are defined in analogue to the classical counterpart, and see how these concepts play together.  Based on these, we finally endow Aharonov's weak value with a statistical interpretation as one realisation of the quantum conditional expectations furnished in our formalism.

\paragraph*{Mathematical Notations Employed}
Throughout this paper, we denote by $\mathbb{K}$ either the real field $\mathbb{R}$ or the complex field $\mathbb{C}$.  Since our primary interest is on quantum mechanics, Hilbert spaces are always assumed to be complex. Conforming to the convention in physical literature, we denote the complex conjugate of a complex number $c \in \mathbb{C}$ by $c^{*}$, and an inner product $\langle \,\cdot\, , \,\cdot\, \rangle$ defined on a complex linear space is anti-linear in its first argument and linear in the second.
For simplicity, we adopt the natural units where we specifically have $\hbar = 1$, unless stated otherwise.

\section{Quantisation and Quasi-classicalisation via QJSDs}\label{sec:two}

For commuting quantum observables, a `trivial' method of quantum and quasi-classical transformation is available, where the former is known as the functional calculus whereas the latter is known as the Born rule.  These maps are both known to be characterised by the joint-spectral measure (JSM) of the observables concerned, and they are understood to be adjoint operations to each other.  On the other hand, the problem becomes non-trivial when non-commuting observables are put in to consideration, primarily due to the lack of the JSM. 

In this section, we first propose a novel approach to the problem of quantum/quasi-classical transformations by introducing the concept of quasi-joint-spectral distributions (QJSDs), which are intended as non-commuting generalisations to the JSMs of commuting observables.  Just as the JSM induces a unique adjoint pair of quantum/quasi-classical transformation for commuting observables, QJSDs induce various adjoint pairs for non-commuting observables.  Specifically, we see that there exists inherent indefiniteness in the possible definition of QJSDs, each leading to different possible transformations, in which the Wigner-Weyl transform belongs as a special case.

\subsection{Preliminary Observations}

As a prelude to our study, we first review some basic facts in quantum theory regarding the spectral theorem of self-adjoint operators, the functional calculus and the Born rule.  In what follows, we consider a finite combination of simultaneously measurable observables, and observe that both the functional calculus and the Born rule can respectively be understood as the trivial realisation of \emph{quantisation} and \emph{quasi-classicalisation}, in the sense that the former allows us to map real functions ({\it i.e.}, classical observables) to self-adjoint Hilbert space operators ({\it i.e.}, quantum observables), and the latter defines a map from density operators ({\it i.e.}, quantum states) to probability distributions ({\it i.e.}, classical states).  We then point out that the functional calculus and the Born rule are \emph{adjoint} notions to each other.  These rather trivial observations shall be the guiding line for our further study in considering the non-trivial problem of quantisation and quasi-classicalisation for the general case involving combination of non-commuting quantum observables.

\subsubsection{Spectral Theorem for Self-adjoint Operators}

A basic but important result of functional analysis ({\it i.e.}, linear algebra for arbitrary-dimensions) regarding self-adjoint operators is the \emph{spectral theorem}, which states that for a self-adjoint operator $A$ on a Hilbert space $\mathcal{H}$, there corresponds a unique \emph{spectral measure} $E_{A}$ such that
\begin{equation}\label{eq:spectral_theorem}
A = \int_{\mathbb{R}} a\ dE_{A}(a)
\end{equation}
holds.  In simple terms, the spectral measure 
\begin{equation}
E_{A}(a) =
\begin{cases}
0 \quad &\text{($a$ is not an eigenvalue of $A$)}\\
P_{a} \quad &\text{($a$ is an eigenvalue of $A$)}
\end{cases}
, \quad a \in \mathbb{R}
\end{equation}
of $A$ is a map from the eigenvalues $a$ of $A$ to the orthogonal projections $P_{a}$ on the corresponding eigenspaces.  For the case where the eigenvalues are all non-degenerate, the spectral measure is simply nothing but $E_{A}(a) = P_{a} = |a\rangle\langle a|$, where $|a\rangle\langle a|$ is the orthogonal projection on the $1$-dimensional subspace of $\mathcal{H}$ spanned by the eigenspace 
corresponding to the eigenvalue $a$.  In this case, the integral \eqref{eq:spectral_theorem} formally reduces to the familiar form
\begin{equation}
A
    = \int_{\mathbb{R}} a |a\rangle\langle a|\ da.
\end{equation}
For the case in which the Hilbert space $\mathcal{H}$ under consideration is moreover finite-dimensional, the spectral theorem is nothing but the eigendecomposition theorem
\begin{equation*}
A = \sum_{i=1}^{n} a_{i} |a_{i}\rangle\langle a_{i}|
\end{equation*}
valid for Hermitian matrices $A$, and the spectral measure $E_{A}(a_{i}) = |a_{i}\rangle\langle a_{i}|$ reduces to the collection of orthogonal projections corresponding to the eigenvectors $a_{i}$ of $A$.  Simply put, spectral theorem is thus a generalisation of the eigendecomposition theorem for the infinite dimensional case, and the spectral measures are in turn the generalisation of orthogonal projections onto the corresponding eigenspaces.  The primary advantage of this generalisation becomes apparent when infinite dimensional Hilbert spaces must be taken into consideration%
\footnote{
A self-adjoint operator defined on infinite dimensional Hilbert spaces may sometimes fail to have any eigenvalues in the sense that $A |\psi\rangle = a |\psi\rangle$ holds for some non-zero vector $|\psi\rangle \in \mathcal{H}$, as most famously exemplified by the position operator $\hat{x}$ and the momentum operator $\hat{p}$ of a free particle.
}.

\paragraph*{Joint-spectral Measures}

Now, suppose one is given an ordered combination
\begin{equation}\label{eq:ordered_combination_of_SA}
\boldsymbol{A} := (A_{1}, \dots, A_{n}), \quad 1 \leq n < \infty
\end{equation}
of a finite number of pairwise strongly commuting%
\footnote{We say that a pair of self-adjoint operators $A$ and $B$ strongly commutes, if and only if they commute in the level of spectral measures.  In a laxer notation, this is to say that
\begin{equation}
E_{A}(a)E_{B}(b) = E_{B}(b)E_{A}(a), \quad a, b \in \mathbb{R}
\end{equation}
holds.  Strictly speaking, strong commutativity is generally stronger than mere commutativity when unbounded operators are concerned, but we do not intend to delve into the intricacies, which are not essential for our discussion.
}
distinct self-adjoint operators ({\it i.e.}, simultaneously measurable quantum observables) on $\mathcal{H}$.  An important fact regarding strongly commuting self-adjoint operators is that, one may uniquely construct the \emph{joint-spectral measure} (JSM) of the combination 
\begin{equation}
E_{\boldsymbol{A}}(\boldsymbol{a}) = \Pi_{i=1}^{n}E_{A_{i}}(a_{i}),\quad \boldsymbol{a} := (a_{1}, \dots, a_{n}) \in \mathbb{R}^{n}
\end{equation}
that fully describes their joint behaviour, where each  $E_{A_{i}}$ is the unique spectral measure corresponding to $A_{i}$, ($1 \leq i \leq n$).  One then trivially has
\begin{equation}
A_{i} = \int_{\mathbb{R}^{n}} a_{i}\ dE_{\boldsymbol{A}}(\boldsymbol{a}), \quad 1 \leq i \leq n,
\end{equation}
if one is to reclaim the original self-adjoint operators.

\subsubsection{Functional Calculus}

An important fact regarding spectral measures is that it induces a map from the space of functions to Hilbert space operators.  Indeed, under the same situation as above, the JSM of the ordered combination \eqref{eq:ordered_combination_of_SA} induces a map that maps a function $f$ defined on $\mathbb{R}^{n}$ to the operator
\begin{equation}\label{eq:functional_calculus_image}
f_{E_{\boldsymbol{A}}} := \int_{\mathbb{R}^{n}} f(\boldsymbol{a})\ dE_{\boldsymbol{A}}(\boldsymbol{a})
\end{equation}
on $\mathcal{H}$.  The map $f \mapsto f_{E_{\boldsymbol{A}}}$ is one realisation of the \emph{functional calculus}, which is a general term that points to a map from functions to operators satisfying certain algebraic properties.  In fact, under some appropriate conditions, one finds that there is a one-to-one correspondence between functional calculi and spectral measures.  In our context, we view the functional calculus as the trivial way to quantise classical observables ({\it i.e.}, real functions) into quantum observables ({\it i.e.}, self-adjoint operators).
In what follows, we may occasionally write the image of the functional calculus by either of the following notations: $f_{E_{\boldsymbol{A}}} = f_{\boldsymbol{A}} = f(\boldsymbol{A}) = f(A_{1}, \dots, A_{n})$.

\subsubsection{Born Rule}

The Born rule is the corner stone of the probabilistic interpretation of quantum measurements, which states that, given a density operator ({\it i.e.}, mixed quantum state) $\rho$ on $\mathcal{H}$ and a combination \eqref{eq:ordered_combination_of_SA} of simultaneously measurable observables, the joint behaviour of the measurement outcomes is described by the \emph{joint-probability distribution}
\begin{equation}\label{eq:born_rule_image}
\rho_{E_{\boldsymbol{A}}}(\boldsymbol{a}) := \mathrm{Tr}[E_{\boldsymbol{A}}(\boldsymbol{a}) \rho]
\end{equation}
defined for the combination of simultaneously measurable observables concerned on the state.  In our context, we view the Born rule as the trivial realisation of quasi-classicalisation of quantum states ({\it i.e.}, density operators) into classical states ({\it i.e.}, probability distributions).  In what follows, we occasionally denote the resulting probability distribution by either of the following notations: $\rho_{E_{\boldsymbol{A}}} = \rho_{\boldsymbol{A}} = \rho_{(A_{1}, \dots, A_{n})}$.

\subsubsection{Adjointness of the Functional Calculus and the Born Rule}

An important observation we point out here that is crucial for our further discussion is that, both quantisation (functional calculus) and quasi-classicalisation (Born rule) are \emph{adjoint} notions.

\paragraph*{Dual Pair}

To see this point, we first prepare some necessary terminologies and notations. Let $L(\mathcal{H})$ denote the space of all bounded linear operators on the Hilbert space $\mathcal{H}$, and let $N(\mathcal{H})$ denote the space of all nuclear operators (or, better known as trace-class operators) on $\mathcal{H}$.  Bounded quantum observables $A \in L(\mathcal{H})$ and quantum states $\rho \in N(\mathcal{H})$ belong to the respective spaces.  On the product space $L(\mathcal{H}) \times N(\mathcal{H})$ is defined a bilinear form
\begin{equation}
\langle X , N \rangle_{Q} := \mathrm{Tr}[XN], \quad X \in L(\mathcal{H}),\ N \in N(\mathcal{H})
\end{equation}
that maps a pair of bounded linear operator and a nuclear operator to the trace of their product.  In mathematics, a triple consisting of a pair of linear spaces $X$, $Y$ and a bilinear form $\langle\,\cdot\, , \,\cdot\,\rangle : X \times Y \to \mathbb{K}$ satisfying the conditions
\begin{equation}
\begin{split}
\forall x \in X \setminus \{0\},\ \exists y \in Y \quad \langle x, y \rangle \neq 0, \\
\forall y \in Y \setminus \{0\},\ \exists x \in X \quad \langle x, y \rangle \neq 0,
\end{split}
\end{equation}
is called a \emph{dual pair}.  The triple $\left( L(\mathcal{H}), N(\mathcal{H}), \langle \,\cdot\, , \,\cdot\, \rangle_{Q} \right)$ is one typical realisation of a dual pair, which are the  familiar tools we use to describes quantum measurements.

On the other hand, let $\mathscr{S}(\mathbb{R}^{n})$, ($1 \leq n < \infty$) denote the $n$-dimensional \emph{Schwartz space}, which is the space of all smooth functions that, even being multiplied by any polynomials after being differentiated arbitrarily many times, they `vanish at infinity'.  Also, let $\mathscr{S}^{\prime}(\mathbb{R}^{n})$ denote the continuous dual of the Schwartz space, which is called the space of \emph{tempered distributions}.  The space of tempered distributions is an extension of the familiar space of density functions: every probability density function is a tempered distribution, while the converse is not always true.  It reveals that the space of density functions (or even that of complex measures) is not sufficient in properly handling quasi-probability distributions for non-commuting observables%
\footnote{
The space of density functions is a proper subset of the space of tempered distributions.  An example for this is the delta distribution, which is a distribution but not a function in the usual sense.
}.
Now, if we allow ourselves some abuse of notation, we may formally treat a tempered distribution $\varphi \in \mathscr{S}^{\prime}(\mathbb{R}^{n})$ as a function $\varphi(\boldsymbol{x})$ on $\mathbb{R}^{n}$, and define the bilinear form
\begin{equation}\label{eq:bilin_classical}
\langle f, \varphi \rangle_{C} := \int_{\mathbb{R}^{n}} f(\boldsymbol{x})\varphi(\boldsymbol{x})\ dm_{n}(\boldsymbol{x}),\quad f \in \mathscr{S}(\mathbb{R}^{n}),\, \varphi \in \mathscr{S}^{\prime}(\mathbb{R}^{n})
\end{equation}
where we have introduced the renormalised Lebesgue-Borel measure
\begin{equation}
dm_{n}(\boldsymbol{x}) := (2\pi)^{-n/2}dx^{n}.
\end{equation}
For brevity, we occasionally write $dm_{1} = dm$ whenever there is no risk for confusion.  This renormalisation is mostly of aesthetic purpose, whose advantage becomes apparent when we introduce the Fourier transformation later in our discussion.  Equipped with the bilinear form, the triple $\left( \mathscr{S}(\mathbb{R}^{n}), \mathscr{S}^{\prime}(\mathbb{R}^{n}), \langle \,\cdot\, , \,\cdot\, \rangle_{C} \right)$ also qualifies as a dual pair that becomes a tool in describing classical measurements.

\paragraph*{Adjointness of the Transformations}

Now, given an ordered combination of simultaneously measurable quantum observables \eqref{eq:ordered_combination_of_SA}, let
\begin{equation}\label{eq:functional_calculus}
\Phi_{E_{\boldsymbol{A}}} : \mathscr{S}(\mathbb{R}^{n}) \to L(\mathcal{H}),\ f \mapsto f_{E_{\boldsymbol{A}}}
\end{equation}
denote the functional calculus of the ordered combination \eqref{eq:ordered_combination_of_SA} of self-adjoint operators defined in \eqref{eq:functional_calculus_image}, and in turn let
\begin{equation}\label{eq:born_rule}
\Phi_{E_{\boldsymbol{A}}}^{\prime}: N(\mathcal{H}) \to \mathscr{S}^{\prime}(\mathbb{R}^{n}),\ \rho \mapsto \rho_{E_{\boldsymbol{A}}}
\end{equation}
denote the Born rule described in \eqref{eq:born_rule_image}.  One then readily observes by the following straightforward computation
\begin{align}
\langle \Phi_{E_{\boldsymbol{A}}}(f), \rho \rangle_{Q}
    &:= \mathrm{Tr}[f_{E_{\boldsymbol{A}}} \rho] \nonumber \\
    &= \int_{\mathbb{R}^{n}} f(\boldsymbol{a})\ d\mathrm{Tr}[E_{\boldsymbol{A}}(\boldsymbol{a}) \rho] \nonumber \\
    &= \int_{\mathbb{R}^{n}} f(\boldsymbol{a}) \rho_{E_{\boldsymbol{A}}}(\boldsymbol{a}) \ dm_{n}(\boldsymbol{a}) \nonumber \\
    &= \langle f, \Phi_{E_{\boldsymbol{A}}}^{\prime}(\rho) \rangle_{C}, \quad f \in \mathscr{S}(\mathbb{R}^{n}),\ \rho \in N(\mathcal{H})
\end{align}
that the functional calculus \eqref{eq:functional_calculus} and the Born rule \eqref{eq:born_rule} are \emph{adjoint maps} to each other.  This relation can be illustrated by the following diagram:
\begin{equation}
\xymatrix@=100pt{
L(\mathcal{H})
	\ar@{<.>}[r]^{\,\text{dual pair}\,}
&  N(\mathcal{H})
	\ar[d]^{\Phi_{E_{\boldsymbol{A}}}^{\prime}}_{\rotatebox{270}{{\footnotesize {\it Quasi-Classicalisation}}}} \\
\mathscr{S}(\mathbb{R}^{n})
    \ar[u]^{\Phi_{E_{\boldsymbol{A}}}}_{\rotatebox{90}{{\footnotesize {\it Quantisation}}}}
    \ar@{<.>}[r]^{\,\text{dual pair}\,}
& \mathscr{S}^{\prime}(\mathbb{R}^{n})
}
\end{equation}
Here, the top row denotes the dual pair of quantum observables and quantum states, whereas the bottom row depicts the classical counterpart.  The left column consists of (quantum and classical) observables, whereas the right column consists of (quantum and classical) states.

\subsection{Quantisation and Quasi-classicalisation via QJSDs}

In the previous subsection, we have reviewed the very basics of the spectral theorem, the functional calculus and the Born rule defined for commuting observables, and have seen that the functional calculus ({\it i.e.}, quantisation) and the born rule ({\it i.e.}, quasi-classicalisation) are adjoint operations to each other.  The next step is to generalise our whole arguments into the case for \emph{non-commuting} observables.

\subsubsection{Introducing Quasi-joint-spectral Distributions (QJSDs)}

The key observation to make here is that, it was the JSM that uniquely gave rise to the adjoint pair of the desired maps.  A straightforward idea for our current problem would be thus to introduce non-commuting analogues to the JSM.

\paragraph*{Strong Commutativity in the Fourier Space}

As a preparation to our further discussion, we review the characterisation for strong commutativity of spectral measures in their Fourier spaces.  Let $A$ be self-adjoint, and let $E_{A}$ be its spectral measure.  We call the operator valued function
\begin{equation}
(\mathscr{F}E_{A})(s) := \int_{\mathbb{R}} e^{-isa}\ dE_{A}(a) = e^{-isA}, \quad s \in \mathbb{R},
\end{equation}
the \emph{Fourier transform} of $E_{A}$.  It is a basic fact of functional calculus that one may characterise the strong commutativity of self-adjoint operators by the Fourier transforms of their spectral measures:  a pair of self-adjoint operators $A$ and $B$ strongly commutes if and only if the Fourier transforms of the respective spectral measures
\begin{equation}\label{eq:strong_commutativity_vs_FT}
e^{itA}e^{isB} = e^{isB}e^{itA}, \quad s,t \in \mathbb{R}
\end{equation}
commute.

\paragraph*{Fourier Transform of JSM}

Let us first compute the Fourier transform of the JSMs.  Given the JSM $E_{\boldsymbol{A}}$ of an ordered combination of strongly commuting self-adjoint operators, let
\begin{align}\label{eq:FT_JSM}
(\mathscr{F}E_{\boldsymbol{A}})(\boldsymbol{s})
    &:= \int_{\mathbb{R}^{n}} e^{-i\langle \boldsymbol{s}, \boldsymbol{a} \rangle}\ dE_{\boldsymbol{A}}(\boldsymbol{a}) \nonumber \\
    &= e^{-i \langle \boldsymbol{s}, \boldsymbol{A} \rangle} \nonumber \\
    &= \Pi_{i=1}^{n} e^{-is_{i}A_{i}}, \quad \boldsymbol{s} \in \mathbb{R}^{n}
\end{align}
denote the \emph{Fourier transform} of $E_{\boldsymbol{A}}$.  Here, $\langle \boldsymbol{s}, \boldsymbol{a} \rangle := \sum_{i=1}^{n} s_{i}a_{i}$ denotes the standard inner product on $\mathbb{R}^{n}$, and $\langle \boldsymbol{s}, \boldsymbol{A} \rangle := \sum_{i=1}^{n} s_{i}A_{i}$.  We also note that the last line of the above equality is due to the iterated application of the Lie-Trotter-Kato product formula.

\paragraph*{Hashed Operators}

From now on, we generalise the use of the notation $\boldsymbol{A}$ so that it may admit ordered combinations of \emph{generally non-commuting} distinct self-adjoint operators.  We then introduce the \emph{hashing}
\begin{multline}\label{def:hashed_operator}
\hat{\#}_{\boldsymbol{A}}(\boldsymbol{s}) := \left\{\text{a `suitable' mixture of the disintegrated } \vphantom{e^{-is_{1}A_{1}}} \right. \\ \left. \text{components of the unitary groups }  e^{-is_{1}A_{1}}, \dots, e^{-is_{n}A_{n}} \right\}, \quad \boldsymbol{s} \in \mathbb{K}^{n}
\end{multline}
of the operators concerned (\emph{hashed operator}).  The term `suitable' is intended to express a mathematical condition as to what qualifies as a reasonable `mixture of the disintegrated components' to meet our purpose, although we do not intend to discuss its precise mathematical definition here.  Examples for the hashing pertaining to the simplest case $\boldsymbol{A} = (A, B)$ are
\begin{align}\label{eq:mixture_examples}
\hat{\#}(s, t) =
    \begin{cases}
        e^{-itB}e^{-isA}, \\
        e^{-isA}e^{-itB}, \\
        \frac{1+\alpha}{2} \cdot e^{-itB}e^{-isA} + \frac{1-\alpha}{2} \cdot e^{-isA}e^{-itB}, \quad \alpha \in \mathbb{C} \\
        \Pi_{k = 1}^{N} e^{-isA/L_{k}}e^{-itB/M_{k}}, \qquad \left(\sum_{k=1}^{N}L_{k}^{-1} = 1,\ \sum_{k=1}^{N}M_{k}^{-1} = 1\right), \\
        \left( e^{-isA/N} e^{-itB/N} \right)^{N}, \\
        e^{-i\overline{(sA + tB)}} = \lim_{N \to \infty} \left( e^{-isA/N} e^{-itB/N} \right)^{N}, \\
        \textit{etc.},
    \end{cases}
\end{align}
or even their adjoints, symmetrisations or convex combinations.  Note that, in general, either or both of the parameters $s$, $t$ can be made to even admits complex numbers.  An example for this case is given by
\begin{equation}\label{def:QJSD_conv_subfamily}
\hat{\#}_{\boldsymbol{A}}^{\kappa}(s, t) = e^{-i\left\langle \frac{1-\kappa}{2},\, s \right\rangle A } e^{-itB} e^{-i\left\langle \frac{1+\kappa}{2},\, s \right\rangle A }, \quad \kappa \in \mathbb{C},\, s \in \mathbb{C},\, t \in \mathbb{R},
\end{equation}
where $\langle \kappa, s \rangle = \kappa_{1}s_{1} + \kappa_{2}s_{2}$ is understood as the standard inner product defined on $\mathbb{R}^{2}$, where $\kappa = \kappa_{1} + i \kappa_{2}$ ($\kappa_{1}, \kappa_{2} \in \mathbb{R}$) and $s = s_{1} + i s_{2}$ ($s_{1}, s_{2} \in \mathbb{R}$) are complex numbers identified as vectors on $\mathbb{R}^{2}$.
One readily realises that the hashed operators \eqref{def:hashed_operator} are, while differing in their representations, unique \emph{if and only if} the self-adjoint operators concerned are all simultaneously measurable.  In that case, it is easy to see that the hashed operators all reduce to the same Fourier transform \eqref{eq:FT_JSM} of the JSM.

\paragraph*{Quasi-joint-spectral Distributions}

Under the same conditions as above, let us choose any hashing $\hat{\#}_{\boldsymbol{A}}$ introduced in \eqref{def:hashed_operator}, and introduce the \emph{quasi-joint-spectral distribution}%
\footnote{
The reason for our choice of the nomination quasi-joint-spectral \emph{distributions}, rather than \emph{measures}, lies in the fact that, in contrast to the JSMs, QJSDs does not necessarily lie in the space of operator valued measures (OVMs).  In fact, we understand them as members of the operator valued distributions (OVDs), which is an operator analogue of generalised functions (distributions).  The space of OVDs is larger than the space of OVMs, and the latter can be embedded into the former.
}
(QJSD) of the ordered pair $\boldsymbol{A}$ defined by its inverse Fourier transform 
\begin{align}
\#_{\boldsymbol{A}}(\boldsymbol{a})
    &:= (\mathscr{F}^{-1}\,\hat{\#}_{\boldsymbol{A}})(\boldsymbol{a}) \nonumber \\
    &:= \int_{\mathbb{K}^{n}} e^{i\langle \boldsymbol{a}, \boldsymbol{s} \rangle} \,\hat{\#}_{\boldsymbol{A}}(\boldsymbol{s})\ dm_{n}(\boldsymbol{s}), \quad \boldsymbol{a} \in \mathbb{K}^{n}.
\end{align}
Due to the bijectivity of the Fourier transformation, to each QJSD corresponds a unique hashed operator, and hence QJSDs are highly non-unique in the case a given ordered combination $\boldsymbol{A}$ is non-commutative.  The QJSD is unique if and only if $\boldsymbol{A}$ admits simultaneous measurability, and in such case, the unique QJSD actually reduces to the JSM itself
\begin{equation}
\#_{\boldsymbol{A}} = E_{\boldsymbol{A}}.
\end{equation}

By construction and the observations made above, one may surmise that the QJSDs serve as generalisations of the JSM to generally non-commuting observables.  Indeed, QJSDs share some of the basic properties one finds in common with the standard JSM.  The primary fact we mention is the normalisation property: the total integration of any QJSD reduces to the identity $\mathrm{Id}$, as one readily finds through the following formal computation
\begin{align}\label{eq:QJSD_normalisation}
\int_{\mathbb{K}^{n}} \,\#_{\boldsymbol{A}}(\boldsymbol{a})\ dm_{n}(\boldsymbol{a})
    &= \int_{\mathbb{K}^{n}} e^{-i \langle \boldsymbol{0}, \boldsymbol{a} \rangle} \,\#_{\boldsymbol{A}}(\boldsymbol{a})\ dm_{n}(\boldsymbol{a}) \nonumber \\
    &= \left( \mathscr{F} \,\#_{\boldsymbol{A}} \right)(\boldsymbol{0}) \nonumber \\
    &= \hat{\#}_{\boldsymbol{A}}(\boldsymbol{0}) = \mathrm{Id}.
\end{align}
In fact, this is actually a corollary to a more stronger property regarding the marginals
\begin{equation}\label{eq:QJSD_marginal}
\int_{\mathbb{K}} \,\#_{\boldsymbol{A}}(\boldsymbol{a})\ dm(a_{k}) = \#_{\boldsymbol{A}_{k}}(a_{1}, \dots, a_{k-1}, a_{k+1}, \dots, a_{n}),
\end{equation}
where $\boldsymbol{A}_{k} := (A_{1}, \dots, A_{k-1}, A_{k+1}, \dots, A_{n})$,  $1 \leq k \leq  n$, denotes the ordered combination of the self-adjoint operators that lacks the $k$th component of the original ordered combination $\boldsymbol{A}$, and $\#_{\boldsymbol{A}_{k}}$ denotes the QJSD of $\boldsymbol{A}_{k}$ defined by 
\begin{equation}
\hat{\#}_{\boldsymbol{A}_{k}}(a_{1}, \dots, a_{k-1}, a_{k+1}, \dots, a_{n}) := \hat{\#}_{\boldsymbol{A}}(a_{1}, \dots, a_{k-1}, 0, a_{k+1}, \dots, a_{n}),
\end{equation}
which corresponds to the hashing constructed by `taking away' all the disintegrated components of the $k$th member $e^{-is_{k}A_{k}}$ from the original hashing $\hat{\#}_{\boldsymbol{A}}$.  To see this, let $H_{k}$ denote the l.~h.~s. of \eqref{eq:QJSD_marginal}. The Fourier transform of $H_{k}$ then reads
\begin{align}
&\mkern-18mu (\mathscr{F}H_{k})(s_{1}, \dots, s_{k-1}, s_{k+1} \dots, s_{n}) \nonumber \\ 
    &= \int_{\mathbb{K}^{n-1}} \sum_{i=1}^{k-1}\sum_{i=k+1}^{n}e^{-is_{i}a_{i}} \left( \int_{\mathbb{K}} \,\#_{\boldsymbol{A}}(\boldsymbol{a})\ dm(a_{k}) \right) \ dm_{n-1}(a_{1}, \dots, a_{k-1}, a_{k+1} \dots, a_{n}) \nonumber \\
    &= \int_{\mathbb{K}^{n}} \sum_{i=1}^{k-1}\sum_{i=k+1}^{n}e^{-is_{i}a_{i}}e^{-i 0 \cdot a_{n}} \,\#_{\boldsymbol{A}}(\boldsymbol{a})\ dm_{n}(\boldsymbol{a}) \nonumber \\
    &= \hat{\#}_{\boldsymbol{A}}(s_{1}, \dots, s_{k-1}, 0, s_{k+1} \dots, s_{n}) \nonumber \\
    &= (\mathscr{F}\#_{\boldsymbol{A}_{k}})(s_{1}, \dots, s_{k-1}, s_{k+1} \dots, s_{n}),
\end{align}
and by the injectivity of the Fourier transformation, one concludes $H_{k} = \#_{\boldsymbol{A}_{k}}$.

\paragraph*{Reclaiming the JSMs}

A straightforward but important corollary to the above property is the following observation.  Let 
\begin{equation}\label{def:order-preserving_combination}
\boldsymbol{B} = (A_{i_{1}}, \dots, A_{i_{k}})
\end{equation}
be an order-preserving ({\it i.e.}, $1 \leq i_{1} < \dots < i_{k} \leq n$) subset of $\boldsymbol{A}$ consisting of $k$ ($1 \leq k \leq n$) numbers of \emph{pairwise strongly commuting} distinct members, and let 
\begin{equation}\label{def:order-preserving_complement}
\boldsymbol{B}^{c} :=  (A_{j_{1}}, \dots, A_{j_{n-k}})
\end{equation}
denote its order-preserving ({\it i.e.}, $1 \leq j_{1} < \dots < j_{n-k} \leq n$)  complement consisting of $n-k$ numbers of the members of $\boldsymbol{A}$ that do not belong to $\boldsymbol{B}$.  Then, an iterated application of  
\eqref{eq:QJSD_marginal} leads to
\begin{equation}\label{eq:reclamation_of_JSM}
E_{\boldsymbol{B}}(\boldsymbol{b}) = \int_{\mathbb{K}^{n-k}} \,\#_{\boldsymbol{A}}(\boldsymbol{a})\ dm_{n-k}(\boldsymbol{b}^{c}),
\end{equation}
where $\boldsymbol{b} := (a_{i_{1}}, \dots, a_{i_{k}})$ and $\boldsymbol{b}^{c} := (a_{j_{1}}, \dots, a_{j_{n-k}})$ denotes the variables corresponding to the respective order-preserving subsets%
\footnote{
Here, we adopt the convention
\begin{equation}
\int_{\mathbb{K}^{n-k}} \,\#_{\boldsymbol{A}}(\boldsymbol{a})\ dm_{n=k}(\boldsymbol{b}^{c}) = \#_{\boldsymbol{A}}(\boldsymbol{a})
\end{equation}
for the case $k=n$.
}.
This implies that, if one `integrate-outs' all the variables corresponding to the complement $\boldsymbol{B}^{c}$, one may reclaim the authentic JSM of $B$.

\subsubsection{Quantisation and Quasi-classicalisation}

Now that we have constructed the QJSDs, which could be understood as non-commutative analogues to the standard JSM, we shall now embark on the construction of quantisation and quasi-classicalisation regarding combination of observables that may fail to be measured simultaneously.

\paragraph*{Quantisation of classical Observables}

For an ordered combination of (generally non-commuting) distinct quantum observables $\boldsymbol{A}$, let $\#_{\boldsymbol{A}}$ be any QJSD of one's choice.  Guided by a straightforward analogy of the functional calculus originally defined for the commutative case, we thus define the map
\begin{equation}\label{def:quantisation}
\Phi_{\#_{\boldsymbol{A}}} : f \mapsto f_{\#_{\boldsymbol{A}}} := \int_{\mathbb{K}^{n}} f(\boldsymbol{a}) \,\#_{\boldsymbol{A}}(\boldsymbol{a})\ dm_{n}(\boldsymbol{a})
\end{equation}
that maps a Schwartz function $f \in \mathscr{S}(\mathbb{K}^{n})$ to a bounded linear operator $f_{\#_{\boldsymbol{A}}} \in L(\mathcal{H})$.  We call the map \eqref{def:quantisation} the \emph{quantisation} pertaining to the QJSD $\#_{\boldsymbol{A}}$, and in turn call the image $\Phi_{\#_{\boldsymbol{A}}}(f)$ the quantisation of $f$.  Occasionally, we denote the quantisation of $f$ by either of the following notations: $f_{\#_{\boldsymbol{A}}} = f(\#_{\boldsymbol{A}})$.  We may sometimes even omit $\boldsymbol{A}$ and write $f_{\#}$, when the observables concerned are obvious from the context.

\paragraph*{Quasi-classicalisation of quantum States}

Conversely, we also allow ourselves to be guided by a straightforward analogy of the Born rule and intend to extend it to the non-commutative case.  We thus define the map
\begin{equation}\label{def:quasi-classicalisation}
\Phi_{\#_{\boldsymbol{A}}}^{\prime} : \rho \mapsto \rho_{\#_{\boldsymbol{A}}}(\boldsymbol{a}) := \mathrm{Tr}\left[\,\#_{\boldsymbol{A}}(\boldsymbol{a}) \rho \right], \quad \boldsymbol{a} \in \mathbb{K}^{n}
\end{equation}
that maps a density operator $\rho \in N(\mathcal{H})$ to a tempered distribution $\rho_{\#_{\boldsymbol{A}}} \in \mathscr{S}^{\prime}(\mathbb{K}^{n})$.  We call the (image $\rho_{\#_{\boldsymbol{A}}}$ of the) map \eqref{def:quasi-classicalisation} the \emph{quasi-classicalisation} (of $\rho$) pertaining to the QJSD $\#_{\boldsymbol{A}}$.   Specifically, for a density operator $\rho \in N(\mathcal{H})$, {\it i.e.}, a positive nuclear operator with the normalisation condition $\mathrm{Tr}[\rho] = 1$, we occasionally call the distribution $\rho_{\#_{\boldsymbol{A}}}$ the \emph{quasi-joint-probability} (QJP) distribution of $\boldsymbol{A}$ on $\rho$ pertaining to the QJSD $\#_{\boldsymbol{A}}$.

\paragraph*{Adjointness of Quantisation and Quasi-classicalisation}

As one may surmise, quantisation \eqref{def:quantisation} and quasi-classicalisation \eqref{def:quasi-classicalisation} are \emph{adjoint} operations to each other, as one may readily check by the formal computation
\begin{align}
\langle \Phi_{\#_{\boldsymbol{A}}}(f), \rho \rangle_{Q}
    &:= \mathrm{Tr}[f_{\#_{\boldsymbol{A}}} \rho] \nonumber \\
    &= \int_{\mathbb{K}^{n}} f(\boldsymbol{a}) \mathrm{Tr}\left[\,\#_{\boldsymbol{A}}(\boldsymbol{a}) \rho \right]\ dm_{n}(\boldsymbol{a}) \nonumber \\
    &= \int_{\mathbb{K}^{n}} f(\boldsymbol{a}) \rho_{\#_{\boldsymbol{A}}}(\boldsymbol{a})\ dm_{n}(\boldsymbol{a}) \nonumber \\
    &= \langle f, \Phi_{\#_{\boldsymbol{A}}}^{\prime}(\rho) \rangle_{C}, \quad f \in \mathscr{S}(\mathbb{R}^{n}),\ \rho \in N(\mathcal{H}).
\end{align}
This relation can be illustrated by the following diagram:
\begin{equation}
\xymatrix@=100pt{
L(\mathcal{H})
	\ar@{<.>}[r]^{\,\text{dual pair}\,}
&  N(\mathcal{H})
	\ar[d]^{\Phi_{\#_{\boldsymbol{A}}}^{\prime}}_{\rotatebox{270}{{\footnotesize {\it Quasi-classicalisation}}}} \\
\mathscr{S}(\mathbb{K}^{n})
    \ar[u]^{\Phi_{\#_{\boldsymbol{A}}}}_{\rotatebox{90}{{\footnotesize {\it Quantisation}}}}
    \ar@{<.>}[r]^{\,\text{dual pair}\,}
& \mathscr{S}^{\prime}(\mathbb{K}^{n})
}
\end{equation}
Note again that, as maps, quantisation \eqref{def:quantisation} and quasi-classicalisation \eqref{def:quasi-classicalisation} are uniquely dictated by the choice of the QJSD $\#_{\boldsymbol{A}}$.  Even for the same classical observable $f \in \mathscr{S}(\mathbb{K}^{n})$, its quantisation generally differs $f_{\#_{\boldsymbol{A}}} \neq f_{\tilde{\#}_{\boldsymbol{A}}}$ given a distinct choice of the QJSD $\#_{\boldsymbol{A}} \neq \tilde{\#}_{\boldsymbol{A}}$, and the same is also true for the quasi-classicalisation of quantum states $\rho \in N(\mathcal{H})$.

\subsection{Quantum/Quasi-classical Representations}

Since quantisation and quasi-classicalisation are adjoint notions to each other, they are different facets of a single entity.  In this sense, we occasionally use the term \emph{quantum/quasi-classical representations} or \emph{transformations} referring to the adjoint pair.
In general, these representations are non-unique, and each of the representations can be specified by the choice of the QJSD, whose indefiniteness originates directly from the non-commutative nature of the observables concerned.

\subsubsection{Transformation of Representations}

We may define transformations of QJSDs in various manners.  In some cases, it may occur that a group of quantisation/quasi-classical representations
could be understood as being \emph{equivalent} to each other in the sense that they can be mutually transformed into one another.
Given an ordered combination $\boldsymbol{A}$ of quantum observables, it is an interesting question to ask ourselves how many QJSDs there are (or in other words, the way of ordering of non-commuting observables) that are essentially distinct to each other up to isomorphisms.

\paragraph*{The Simplest Case}

The simplest case for this is when two QJSDs, while being distinct in their form as hashed operators, are identical.  Needless to say, this is trivially always the case when all the members of the ordered combination $\boldsymbol{A}$ pairwise strongly commute.  As an example of less trivial cases, let $\boldsymbol{X} = (Q,P)$ be an ordered pair of self-adjoint operators such that the members satisfy the Weyl representation
\begin{equation}\label{def:Weyl_rep_CCR}
e^{-isQ}e^{-itP} = e^{-ist}e^{-itP}e^{-isQ}, \quad s,\, t \in \mathbb{R}
\end{equation}
of the canonical commutation relation (CCR).
Then, by a simple observation, one verifies that the hashed operators of the form
\begin{align}\label{eq:Wigner-Weyl_equivalence}
e^{-i\frac{t}{2}P}e^{-isQ}e^{-i\frac{t}{2}P} 
    &= e^{-i\frac{s}{3}Q}e^{-i\frac{t}{2}P}e^{-i\frac{s}{3}Q}e^{-i\frac{t}{2}P}e^{-i\frac{s}{3}Q} \nonumber \\
    &= e^{-i\frac{t}{4}P}e^{-i\frac{s}{3}Q}e^{-i\frac{t}{4}P}e^{-i\frac{s}{3}Q}e^{-i\frac{t}{4}P}e^{-i\frac{s}{3}Q}e^{-i\frac{t}{4}P} \nonumber \\
    &= \cdots \nonumber \\
    &= e^{-i\overline{(sQ + tP)}} \nonumber \\
    &=  \cdots \nonumber \\
    &= e^{-i\frac{s}{4}Q}e^{-i\frac{t}{3}P}e^{-i\frac{s}{4}Q}e^{-i\frac{t}{3}P}e^{-i\frac{s}{4}Q}e^{-i\frac{t}{3}P}e^{-i\frac{s}{4}Q} \nonumber \\
    &= e^{-i\frac{t}{3}P}e^{-i\frac{s}{2}Q}e^{-i\frac{t}{3}P}e^{-i\frac{s}{2}Q}e^{-i\frac{t}{3}P} \nonumber \\
    &= e^{-i\frac{s}{2}Q}e^{-itP}e^{-i\frac{s}{2}Q}, \quad s,\ t \in \mathbb{R},
\end{align}
are all identical.  Here, the equality in the center is due to the product formula proven by Lie-Trotter-Kato, and the overline on the operator $sQ + tP$ denotes its self-adjoint extension.  Although they differ in their representation as mixtures, their corresponding QJSDs are identical, and thus all the quantisation/quasi-classical representations induced could be naturally understood as being equivalent.

\paragraph*{Affine Transfromations}

As another example, let $T: \mathbb{K}^{n} \to \mathbb{K}^{n}$ be a linear map, and consider the affine map
\begin{equation}\label{def:affine_transformation}
T_{\boldsymbol{b}} : \boldsymbol{a} \mapsto T\boldsymbol{a} + \boldsymbol{b}
\end{equation}
defined for $\boldsymbol{b} \in \mathbb{K}^{n}$.  We then introduce the \emph{affine transform} of a QJSD $\#_{\boldsymbol{A}}$ with respect to the affine map \eqref{def:affine_transformation} by
\begin{equation}\label{def:QJSD_affine_transform}
\left(T_{\boldsymbol{b}}\#_{\boldsymbol{A}}\right)(\boldsymbol{a}) := \#_{\boldsymbol{A}}\left(T_{\boldsymbol{b}}^{-1}\boldsymbol{a}\right), \quad \boldsymbol{a} \in \mathbb{K}^{n}.
\end{equation}
If $T$ is a bijection ({\it i.e.}, $\det T \neq 0$), both $T_{\boldsymbol{b}}\#_{\boldsymbol{A}}$ and $\#_{\boldsymbol{A}}$ are invertible to one another and thus essentially contain the same information of the combination $\boldsymbol{A}$ of the observables concerned.  It is to be noted that the affine transform of a QJSD is generally not a member of the QJSDs.  Indeed, while the total integration of \eqref{def:QJSD_affine_transform} reduces to the identity $\mathrm{Id}$ by definition, it most importantly fails to satisfy the marginal properties \eqref{eq:QJSD_marginal} in general.  
Even so, affine transforms of a QJSD give rise to adjoint pairs of quantisation and quasi-classicalisation in an extended sense, which are respectively defined by
\begin{equation}
\Phi_{\left(T_{\boldsymbol{b}}\#_{\boldsymbol{A}}\right)} : f \mapsto f_{\left(T_{\boldsymbol{b}}\#_{\boldsymbol{A}}\right)} := \int_{\mathbb{K}^{n}} f(\boldsymbol{a}) \left(T_{\boldsymbol{b}}\#_{\boldsymbol{A}}\right)(\boldsymbol{a}) \ dm_{n}(\boldsymbol{a})
\end{equation}
and
\begin{equation}
\Phi_{\left(T_{\boldsymbol{b}}\#_{\boldsymbol{A}}\right)}^{\prime} : \rho \mapsto \rho_{\left(T_{\boldsymbol{b}}\#_{\boldsymbol{A}}\right)}(\boldsymbol{a}) := \mathrm{Tr}[\left(T_{\boldsymbol{b}}\#_{\boldsymbol{A}}\right)(\boldsymbol{a})\rho].
\end{equation}
The adjointness of these operations
\begin{align}
\langle f_{\left(T_{\boldsymbol{b}}\#_{\boldsymbol{A}}\right)}, \rho \rangle_{Q}
    &= \langle f, \rho_{\left(T_{\boldsymbol{b}}\#_{\boldsymbol{A}}\right)} \rangle_{C}
\end{align}
can be readily confirmed by a simple computation.  To see how the quantum/quasi-classical representations corresponding to affine transforms relate to the original representation, we first observe
\begin{align}
f_{\left(T_{\boldsymbol{b}}\#_{\boldsymbol{A}}\right)}
    &:= \int_{\mathbb{K}^{n}} f(\boldsymbol{a}) \left(T_{\boldsymbol{b}}\#_{\boldsymbol{A}}\right)(\boldsymbol{a}) \ dm_{n}(\boldsymbol{a})\nonumber \\
    &= \int_{\mathbb{K}^{n}} f(T_{\boldsymbol{b}}\boldsymbol{a}) \#_{\boldsymbol{A}}(\boldsymbol{a}) \ dm_{n}(\boldsymbol{a}) \nonumber \\
    &= (f \circ T_{\boldsymbol{b}})_{\#_{\boldsymbol{A}}} \nonumber \\
    &=: (T_{\boldsymbol{b}}^{*}f )_{\#_{\boldsymbol{A}}} 
\end{align}
and
\begin{align}
\rho_{\left(T_{\boldsymbol{b}}\#_{\boldsymbol{A}}\right)}(\boldsymbol{a})
    &= \rho_{\#_{\boldsymbol{A}}} \left( T_{\boldsymbol{b}}^{-1}\boldsymbol{a} \right), \nonumber \\
    &=: T_{\boldsymbol{b}*}\rho_{\#_{\boldsymbol{A}}}
\end{align}
where
\begin{align}
T_{\boldsymbol{b}}^{*}f &:= f \circ T_{\boldsymbol{b}}, \\
T_{\boldsymbol{b}*}\rho &:= \rho \circ T_{\boldsymbol{b}}^{-1},
\end{align}
respectively denote the pullback of a function $f$ and the pushforward of a distribution $\rho$ by the affine map $T_{\boldsymbol{b}}$.
The relation can thus be illustrated by the following diagram
\begin{equation*}
\xymatrix@=40pt{
L(\mathcal{H})
    \ar@{<.>}[rrr]^{\,\text{dual pair}\,}
&
&
& N(\mathcal{H})
    \ar[dl]_{\Phi_{\#_{\boldsymbol{A}}}^{\prime}}
    \ar[dd]^{\Phi_{\left(T_{\boldsymbol{b}}\#_{\boldsymbol{A}}\right)}^{\prime}}
\\
&\mathscr{S}(\mathbb{K}^{n})
    \ar@{<.>}[r]^{\,\text{dual pair}\,}
    \ar[ul]_{\Phi_{\#_{\boldsymbol{A}}}}
& \mathscr{S}^{\prime}(\mathbb{K}^{n})
    \ar[dr]^{T_{\boldsymbol{b}*}}
&
\\
 \mathscr{S}(\mathbb{K}^{n})
    \ar[uu]^{\Phi_{\left(T_{\boldsymbol{b}}\#_{\boldsymbol{A}}\right)}}
    \ar@{<.>}[rrr]^{\,\text{dual pair}\,}
    \ar[ur]^{T_{\boldsymbol{b}}^{*}}
&
& 
& \mathscr{S}^{\prime}(\mathbb{K}^{n})
}
\end{equation*}

As a simple concrete example, we consider the simplest case $\boldsymbol{A} = (A,B)$.  Below, we see that all the members $\#_{\boldsymbol{A}}^{\kappa}$ of the subfamily of the QJSDs of the form \eqref{def:QJSD_conv_subfamily} are linear transforms of $\#_{\boldsymbol{A}}^{i}$ for the specific choice $\kappa = i$.  To see this, consider the matrix
\begin{align}
\tilde{T}_{\kappa} := \left(
    \begin{array}{cc}
    1 & \kappa_{1} \\
    0 & \kappa_{2}
    \end{array}
    \right)
\end{align}
defined for each $\kappa = \kappa_{1} + i \kappa_{2}$, ($\kappa_{1}, \kappa_{2} \in \mathbb{R}$), and the linear transformation $T_{\kappa} := \tilde{T}_{\kappa} \times \mathrm{Id}$ on $\mathbb{C} \times \mathbb{R}$ defined by $T_{\kappa} (a,\, b) := (\tilde{T}_{\kappa}a,\, b)$.  Then, a simple computation yields
\begin{align}
\mathscr{F} \left( T_{\kappa}\,\#_{\boldsymbol{A}}^{i} \right)(s,t)
    &= \int_{\mathbb{C} \times \mathbb{R}} e^{-i\langle s, T_{\kappa}a \rangle}e^{-itb} \,\#_{\boldsymbol{A}}^{i}(a,b)\ dm_{2}(a,b) \nonumber \\
    &= \hat{\#}_{\boldsymbol{A}}^{i}(T_{\kappa}^{\prime}s,t) \nonumber \\
    &= \mathscr{F}\#_{\boldsymbol{A}}^{\kappa}(s,t),
\end{align}
where $T_{\kappa}^{\prime}$ denotes the adjoint matrix of $T_{\kappa}$.  We thus conclude
\begin{equation}
T_{\kappa}\,\#_{\boldsymbol{A}}^{i} = \#_{\boldsymbol{A}}^{\kappa}.
\end{equation}
Since $\det T_{\kappa} = \kappa_{2}$, it is straightforward to see that all the members $\#_{\boldsymbol{A}}^{\kappa}$ of the QJSDs for the choice $\mathrm{Im}\, \kappa \neq 0$ are equivalent to one another by linear transformations.

\paragraph*{Convolutions}

As another important class of transformations, we consider the \emph{convolution}
\begin{equation}\label{def:conv_QJSD_func}
(h \ast \#_{\boldsymbol{A}})(\boldsymbol{a}) := \int_{\mathbb{K}^{n}} h(\boldsymbol{a} - \boldsymbol{a}^{\prime})\,\#_{\boldsymbol{A}}(\boldsymbol{a}^{\prime})\ dm_{k}(\boldsymbol{a}^{\prime})
\end{equation}
of a QJSD $\#_{\boldsymbol{A}}$ and a function $h$ with the total integration of unity.  The Fourier transform of the convolution reads
\begin{align}\label{eq:conv_QJSD_func_as_average}
\mathscr{F} (h \ast \#_{\boldsymbol{A}})(\boldsymbol{s})
    &= \hat{h}(\boldsymbol{s}) \,\hat{\#}_{\boldsymbol{A}}(\boldsymbol{s}) \nonumber \\
    &= \int_{\mathbb{K}^{n}} h(\boldsymbol{k}) e^{-i\langle \boldsymbol{s},\,\boldsymbol{k}\rangle} \,\hat{\#}_{\boldsymbol{A}}(\boldsymbol{s})\ dm_{n}(\boldsymbol{k}) \nonumber \\
    &= \int_{\mathbb{K}^{n}} h(\boldsymbol{k}) \,\hat{\#}_{\boldsymbol{A} + \boldsymbol{k}}(\boldsymbol{s})\ dm_{n}(\boldsymbol{k}),
\end{align}
where $\boldsymbol{A} + \boldsymbol{k} := (A_{1} + k_{1} \cdot \mathrm{Id}, \dots, A_{n} + k_{n} \cdot \mathrm{Id})$ denotes the ordered combination of the normal operators defined as the parallel translation of $\boldsymbol{A}$ towards the direction $\boldsymbol{k} \in \mathbb{K}^{n}$.  From this, the convolution
\begin{equation}\label{eq:conv_as_weighted_average}
h \ast \#_{\boldsymbol{A}} = \int_{\mathbb{K}^{n}} h(\boldsymbol{k})\, \#_{\boldsymbol{A} + \boldsymbol{k}}\ dm_{n}(\boldsymbol{k})
\end{equation}
could be understood as the `weighted average' of the family of QJSDs $\#_{\boldsymbol{A} + \boldsymbol{k}}$ of the ordered combination $\boldsymbol{A} + \boldsymbol{k}$ of normal operators, or equivalently, the parallel translation
\begin{equation}
\#_{\boldsymbol{A} + \boldsymbol{k}} = \tau_{\boldsymbol{k}}\#_{\boldsymbol{A}}, \quad (\tau_{\boldsymbol{k}} \boldsymbol{a} := \boldsymbol{a} + \boldsymbol{k})
\end{equation}
of the original QJSD $\#_{\boldsymbol{A}}$, with respect to the `weight function' $h$.

It is important to note that, in general, the convolution \eqref{def:conv_QJSD_func} itself is not necessarily a QJSD of $\boldsymbol{A}$.   Indeed, consider the most extreme case where all the members of $\boldsymbol{A}$ pairwise strongly commute.  In such case, the unique QJSD of $\boldsymbol{A}$ is the JSM $E_{\boldsymbol{A}}$, so in order for the convolution $h \ast E_{\boldsymbol{A}} = E_{\boldsymbol{A}}$ to be the unique QJSD of $\boldsymbol{A}$, we must have $h = \delta_{\boldsymbol{0}} \Leftrightarrow \hat{h} = 1$.  In a more general setting, a necessary condition for $h \ast \#_{\boldsymbol{A}}$ to satisfy the marginal condition \eqref{eq:reclamation_of_JSM} is given by%
\footnote{
Here, we adopt the convention
\begin{equation}
\int_{\mathbb{K}^{n-k}} h(\boldsymbol{a})\ dm_{n-k}(\boldsymbol{b}^{c}) = h(\boldsymbol{a})
\end{equation}
for the case $k=n$.
}
\begin{equation}\label{eq:condition_conv_QJSD}
\int_{\mathbb{K}^{n-k}} h(\boldsymbol{a})\ dm_{n-k}(\boldsymbol{b}^{c}) = \delta_{\boldsymbol{0}}(\boldsymbol{b}),
\end{equation}
where $\boldsymbol{B}$ is an order-preserving subset \eqref{def:order-preserving_combination} of $\boldsymbol{A}$ consisting of $k$ ($1 \leq k \leq n$) numbers of its pairwise strongly commuting distinct members, $\boldsymbol{B}^{c}$ the order-preserving complement \eqref{def:order-preserving_complement} of $\boldsymbol{B}$, and $\boldsymbol{b} \in \mathbb{K}^{k}$, $\boldsymbol{b}^{c} \in \mathbb{K}^{n-k}$ are their corresponding variables.
Here, $\delta_{\boldsymbol{a}}$, denotes the \emph{delta distribution} centred at $\boldsymbol{a} \in \mathbb{K}^{n}$, which is a generalised function symbolically defined as
\begin{equation}
\delta_{\boldsymbol{a}}(\boldsymbol{x})=
\begin{cases}
\infty, \quad &(\boldsymbol{x} = \boldsymbol{a}) \\
0, \quad &(\boldsymbol{x} \neq \boldsymbol{a})
\end{cases}
\end{equation}
in the usual manner.

On the other hand, the convolution still gives rise to the adjoint pair of representations in an extended sense in a similar manner to the affine transforms of a QJSD.  In order to see its relation to the original representation, we first observe that
\begin{align}
f_{(h\ast\#_{\boldsymbol{A}})}
    &:= \int_{\mathbb{K}^{n}} f(\boldsymbol{a}) (h \ast \#_{\boldsymbol{A}})(\boldsymbol{a})\ dm_{n}(\boldsymbol{a}) \nonumber \\
    &= \int_{\mathbb{K}^{n}} (\tilde{h} \ast f)(\boldsymbol{a}) \#_{\boldsymbol{A}}(\boldsymbol{a})\ dm_{n}(\boldsymbol{a}) \nonumber \\
    &=: (\tilde{h} \ast f)_{\#_{\boldsymbol{A}}},
\end{align}
where $\tilde{h}(\boldsymbol{a}) := h(-\boldsymbol{a})$ denotes the transpose of $h$.  One also finds
\begin{align}\label{eq:convolution_transformation_classicalisation}
\rho_{(h \ast \#_{\boldsymbol{A}})} = h \ast \rho_{\#_{\boldsymbol{A}}}.
\end{align}
The adjointness of these operations
\begin{align}
\left\langle f_{(h\ast\#_{\boldsymbol{A}})}, \rho \right\rangle_{Q}
    &= \left\langle f, \rho_{(h \ast \#_{\boldsymbol{A}})} \right\rangle_{C}
\end{align}
can be readily confirmed by a simple computation.  
The diagram
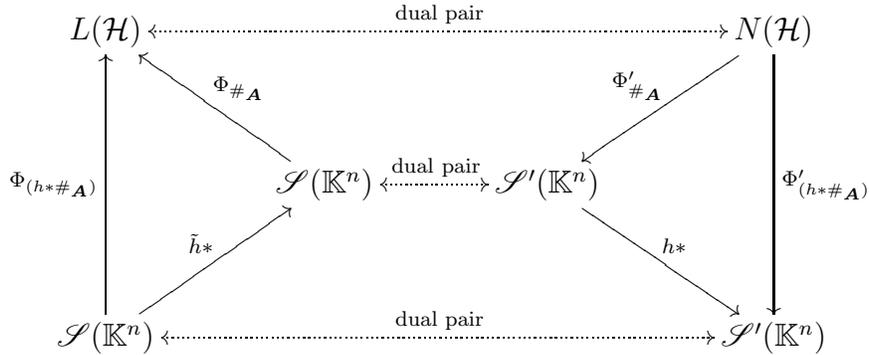
\begin{equation*}
\xymatrix@=40pt{
L(\mathcal{H})
    \ar@{<.>}[rrr]^{\,\text{dual pair}\,}
&
&
& N(\mathcal{H})
    \ar[dl]_{\Phi_{\#_{\boldsymbol{A}}}^{\prime}}
    \ar[dd]^{\Phi_{(h \ast \#_{\boldsymbol{A}})}^{\prime}}
\\
&\mathscr{S}(\mathbb{K}^{n})
    \ar@{<.>}[r]^{\,\text{dual pair}\,}
    \ar[ul]_{\Phi_{\#_{\boldsymbol{A}}}}
& \mathscr{S}^{\prime}(\mathbb{K}^{n})
    \ar[dr]^{h \ast}
&
\\
 \mathscr{S}(\mathbb{K}^{n})
    \ar[uu]^{\Phi_{(h \ast \#_{\boldsymbol{A}})}}
    \ar@{<.>}[rrr]^{\,\text{dual pair}\,}
    \ar[ur]^{\tilde{h} \ast }
&
& 
& \mathscr{S}^{\prime}(\mathbb{K}^{n})
}
\end{equation*}
gives a visual summary as to how the quantum/quasi-classical representation corresponding to the convolution of the QJSD $\#_{\boldsymbol{A}}$ with the function $h$ relates to the original representation.  One readily sees that the two representations corresponding to $\#_{\boldsymbol{A}}$ and $h \ast \#_{\boldsymbol{A}}$ are equivalent if the map $h \ast : f \mapsto h \ast f$ is a bijection.

\subsubsection{Faithfulness of the Representations}

We say that the quantum/quasi-classical representation pertaining to a given QJSD $\#_{\boldsymbol{A}}$ is \emph{faithful}, if either of the following equivalent%
\footnote{
The proof for the equivalence of the conditions can be carried out by applying the Hahn-Banach theorem on locally convex spaces.
}
conditions are met:
\begin{enumerate}
\item The quantisation $\Phi_{\#_{\boldsymbol{A}}}$ has a dense range, {\it i.e.},
\begin{equation}
\overline{ \mathrm{ran}\, \Phi_{\#_{\boldsymbol{A}}}} = L(\mathcal{H}).
\end{equation}
\item The quasi-classicalisation $\Phi_{\#_{\boldsymbol{A}}}^{\prime}$ is injective.
\end{enumerate}
In physical terms, this is to say that every quantum state $\rho$ can be uniquely distinguished by the resulting QJP distribution $\rho_{\#_{\boldsymbol{A}}}$, and that every quantum operator $X = \lim f_{i}(\#_{\boldsymbol{A}})$ can be represented by limits of quantised operators.  In general, the larger the number of observables belonging to the ordered combination $\boldsymbol{A}$ becomes, the closer the representation approaches to faithfulness.

\subsubsection{Realness of the Representations}

Among the various candidates of representations, `real' representations are oftentimes prized.  To state the precise definition, we first need some preparations.
Given a QJSD $\#_{\boldsymbol{A}}$, the \emph{conjugate} of $\#_{\boldsymbol{A}}$ is formally defined by
\begin{equation}
\#_{\boldsymbol{A}}^{*}(\boldsymbol{a}) := \#_{\boldsymbol{A}}(\boldsymbol{a})^{*}, \quad \boldsymbol{a} \in \mathbb{K}^{n},
\end{equation}
where the asterisk on the r.~h.~s. denotes the adjoint of $\#_{\boldsymbol{A}}(\boldsymbol{a})$.  The Fourier transform of the conjugate of a QJSD reads
\begin{align}
(\mathscr{F}\#_{\boldsymbol{A}}^{*})(\boldsymbol{s})
    &:= \int_{\mathbb{K}^{n}} e^{-i \langle \boldsymbol{s}, \boldsymbol{a} \rangle} \#_{\boldsymbol{A}}(\boldsymbol{a})^{*}\ dm_{n}(\boldsymbol{a}) \nonumber \\
    &= \left( \int_{\mathbb{K}^{n}} e^{-i \langle -\boldsymbol{s}, \boldsymbol{a} \rangle} \#_{\boldsymbol{A}}(\boldsymbol{a})\ dm_{n}(\boldsymbol{a}) \right)^{*} \nonumber \\
    &= \hat{\#}_{\boldsymbol{A}}(-\boldsymbol{s})^{*} \nonumber \\
    &=: \hat{\#}_{\boldsymbol{A}}^{\dagger}(\boldsymbol{s}),
\end{align}
where we have introduced the \emph{involution} $\hat{\#}_{\boldsymbol{A}}^{\dagger}$ of the hashed operator $\hat{\#}_{\boldsymbol{A}}$ in the last equality. It is not difficult to see that involutions of hashed operators are again hashed operators, hence a conjugate of a QJSD is a QJSD.

We say that a QJSD $\#_{\boldsymbol{A}}$ is \emph{real} if $\#_{\boldsymbol{A}}^{*} = \#_{\boldsymbol{A}}$ holds.  By the bijectivity of the Fourier transformation, a QJSD is real if and only if its Fourier transform ({\it i.e.}, the corresponding hashed operator) is a self-involution.  Examples of real QJSDs of $\boldsymbol{A} = (A,B)$ are those whose corresponding hashed operators read
\begin{align}
\hat{\#}(s,t) =
\begin{cases}
\frac{1}{2} \cdot (e^{-isB}e^{-isA} + e^{-isA}e^{-itB}) \\
e^{-i\frac{s}{2}A}e^{-itB}e^{-i\frac{s}{2}A} \\
e^{-i\frac{t}{2}B}e^{-isA}e^{-i\frac{t}{2}B} \\
e^{-i\frac{s}{3}A}e^{-i\frac{t}{2}B}e^{-i\frac{s}{3}A}e^{-i\frac{t}{2}B}e^{-i\frac{s}{3}A} \nonumber \\
e^{-i\frac{t}{4}B}e^{-i\frac{s}{3}A}e^{-i\frac{t}{4}B}e^{-i\frac{s}{3}A}e^{-i\frac{t}{4}B}e^{-i\frac{s}{3}A}e^{-i\frac{t}{4}B} \\
e^{-i\overline{(sA + tB)}} \\
\mathit{etc.}
\end{cases}
\quad s,\ t \in \mathbb{R},
\end{align}
were the overline on the operator $sA + tB$ denotes its unique self-adjoint extension.  Colloquially speaking, a QJSD is real if its corresponding hashed operator is `symmetric' in its form.  We call the quantum/quasi-classical representation real, if the corresponding QJSD is real.  Real representations have the following convenient properties:
\begin{enumerate}
\item The quantisation $f_{\#_{\boldsymbol{A}}}$ of a real function $f \in \mathscr{S}(\mathbb{K}^{n})$ is always self-adjoint.
\item The quasi-classicalisation $\rho_{\#_{\boldsymbol{A}}}$ of a density operator $\rho \in N(\mathcal{H})$ is always real.
\end{enumerate}
Real representations thus have a formal advantage of taking a classical observables into self-adjoint operators, and a quantum state into real QJP distribution, which some may find favourable above non-real representations.

\subsubsection{Relation to some prior Works}

The study on quantum-classical transformation has a long history.  In this passage, we investigate the relation of the formalism of QJSDs presented in this paper to some of the prior works on this topic.  In this passage, we are specifically interested in the choice $\boldsymbol{X} = (Q,P)$, where $Q$ and $P$ are operators satisfying the Weyl representation \eqref{def:Weyl_rep_CCR} of the CCR. 

\paragraph*{The Theory of Wigner and Weyl}

We first point out that the Weyl map and the Wigner map are respectively the quantisation $\Phi_{\#_{\boldsymbol{X}}}$ and the quasi-classicalisation $\Phi_{\#_{\boldsymbol{X}}}^{\prime}$ pertaining to the QJSD $W_{\boldsymbol{X}}$ of $\boldsymbol{X}$ corresponding to the hashed operator of the form
\begin{equation}\label{eq:Wigner-Weyl_Hash}
\hat{W}_{\boldsymbol{X}}(s,t)
    := e^{-i\overline{(sQ + tP)}}.
\end{equation}
In fact, since the hashed operator \eqref{eq:Wigner-Weyl_Hash} is equivalent to any choice of the form \eqref{eq:Wigner-Weyl_equivalence}, the Wigner-Weyl transformation is the quantum/quasi-classical representation pertaining to any of the choices.  It is easy to see from the self-involutive form of the hashed operator \eqref{eq:Wigner-Weyl_Hash} that the Wigner-Weyl transformation is real.  It is widely known that the Wigner map is injective, which is equivalent for its adjoint map ({\it i.e.}, the Weyl map) to have a dense range:  in the terminology of this paper, the Wigner-Weyl transformation is faithful.

In order to confirm the claim, we first compute the quantisation of functions with respect to the QJSD corresponding to \eqref{eq:Wigner-Weyl_Hash}.  The quantisation of $f \in \mathscr{S}(\mathbb{R}^{2})$ reads
\begin{align}
f_{W_{\boldsymbol{X}}}
    &:= \int_{\mathbb{R}^{2}} f(q,p)\ W_{\boldsymbol{X}}(q,p) dm_{2}(q,p) \nonumber \\
    &= \int_{\mathbb{R}^{4}} f(q,p) e^{i(qs + pt)}\ \hat{W}_{\boldsymbol{X}}(s,t) dm_{2}(q,p,s,t) \nonumber \\ 
    &= \int_{\mathbb{R}^{4}} f(q,p) e^{-is(Q - q)}e^{-it(P - p)} e^{ist/2} dm_{2}(q,p,s,t) \nonumber \\
    &=: W_{\boldsymbol{X}}(f),
\end{align}
where the last equality is precisely the definition of the Weyl quantisation of $f$.  Here, note that we have used the relation
\begin{equation}
e^{-itP/2}e^{-isQ}e^{-itP/2} = e^{ist/2}e^{-isQ}e^{-itP}
\end{equation}
in order to obtain the third equality.  As for the quasi-classical representation of a quantum state, we assume without loss of generality that $\mathcal{H} = L^{2}(\mathbb{R})$, and that $Q = \hat{x}$, $P = \hat{p}$ are respectively the familiar position and momentum operators.  For better readability, we moreover restrict ourselves to the case that $\rho = |\psi\rangle\langle \psi|$ for some wave-function $\psi \in L^{2}(\mathbb{R})$.  Then, the Fourier transform of $\rho_{W_{\boldsymbol{X}}}$ reads
\begin{align}
\left( \mathscr{F}\rho_{W_{\boldsymbol{X}}} \right)(s,t)
    &= \mathrm{Tr}\left[ \hat{W}_{\boldsymbol{X}}(s,t) \rho \right] \nonumber \\
    &= \langle e^{itP/2} \psi, e^{-isQ} e^{-itP/2} \psi\rangle \nonumber \\
    &= \int_{\mathbb{R}} e^{-isq} \psi^{*}(q+t/2)\psi(q-t/2)\ dm(x) \nonumber \nonumber \\
    &= \int_{\mathbb{R}} e^{-i(sq + tp)} \left( \int_{\mathbb{R}} \psi^{*}(q+t/2)\psi(q-t/2) e^{ipt}\ dm(t) \right) dm_{2}(q,p) \nonumber \\
    &= \left( \mathscr{F}W_{\boldsymbol{X}}^{\rho} \right)(s,t),
\end{align}
where
\begin{equation}
W_{\boldsymbol{X}}^{\rho}(q,p) := \int_{\mathbb{R}} \psi^{*}(q+t/2)\psi(q-t/2) e^{ipt}\ dm(t)
\end{equation}
is the Wigner function of the wave function $\psi \in L^{2}(\mathbb{R})$.  Due to the injectivity of the Fourier transformation, we conclude $\rho_{W_{\boldsymbol{X}}} = W_{\boldsymbol{X}}^{\rho}$.  The proof for the general case goes essentially the same.

\paragraph*{QSJDs generated by Convolution}

We next consider the problem of generating a family of QJSDs of $\boldsymbol{X}$ by means of the convolution of $W_{\boldsymbol{X}}$  with functions.  This setting has been previously examined by Cohen et al. \cite{Cohen_1987} with the intention of constructing a family of generalised phase space distribution functions.  For the convolution $h \ast W_{\boldsymbol{X}}$ to be a QJSD of $\boldsymbol{X}$, one sees from the result \eqref{eq:condition_conv_QJSD} that the condition $\hat{h}(0,t) = \hat{h}(s,0) = 1$, $s,t \in \mathbb{R}$ is necessary.  Specifically, we demonstrate below that, under the assumption that the Fourier transform
\begin{equation}
\hat{h}(s,t) = \hat{g}\left(st/2\right), \quad s,\,t \in \mathbb{R}
\end{equation}
of the function $h$ can be represented by a function $g : \mathbb{R} \to \mathbb{R}$ with the total integration of unity, the convolution \eqref{eq:conv_as_weighted_average} becomes a QJSD of the pair $\boldsymbol{X}$.  Indeed, observe that since 
\begin{equation}
\mathscr{F} (h \ast W_{\boldsymbol{X}})(s,t)
    = \int_{\mathbb{R}} g(\kappa) e^{-i\kappa st/2} \,\hat{W}_{\boldsymbol{X}}(s,t)\ dm(\kappa),
\end{equation}
we have
\begin{equation}\label{eq:conv_QJSD_special}
h \ast W_{\boldsymbol{X}} = \int_{\mathbb{R}} g(\kappa) \,\#_{\boldsymbol{X}}^{\kappa}(s,t)\ dm(\kappa),
\end{equation}
where we have used the parametrised family $\#_{\boldsymbol{X}}^{\kappa}$, $\kappa \in \mathbb{R}$, of the QJSD of $\boldsymbol{X}$ corresponding to the hashed operators of the form
\begin{align}
\hat{\#}_{\boldsymbol{X}}^{\kappa}(s,t)
    &:= e^{-i\kappa st/2} \,\hat{W}_{\boldsymbol{X}}(s,t) \nonumber \\
    &= e^{-i \frac{1-\kappa}{2} sQ} e^{-itP} e^{-i \frac{1+\kappa}{2} sQ},
\end{align}
which was originally introduced in \eqref{def:QJSD_conv_subfamily} for general $\boldsymbol{A} = (A,B)$.  Hence, as the `weighted average' of the parametrised families of the QJSDs of $\boldsymbol{X}$, the convolution \eqref{eq:conv_QJSD_special} itself is a QJSD of $\boldsymbol{X}$.  Each choice of the function $g$ yields distinct representation.  Among the well-known representations are those proposed by:
\begin{enumerate}
\item Weyl \cite{Weyl_1927} and Wigner \cite{Wigner_1932}
\begin{equation}\label{eq:QJP_Weyl_Wigner}
g(k) = \delta_{0}(k) \quad \Leftrightarrow \quad \hat{g}(\omega) = 1
\end{equation}
\item Kirkwood \cite{Kirkwood_1933} and Dirac \cite{Dirac_1945}
\begin{equation}\label{eq:QJP_Kirkwood_Dirac}
g(k) = \delta_{\pm 1}(k) \quad \Leftrightarrow \quad \hat{g}(\omega) = e^{\mp i\omega}
\end{equation}
\item Margenau and Hill \cite{Margenau-Hill_1961}
\begin{equation}\label{eq:QJP_Margenau_Hill}
g(k) = \frac{\left( \delta_{-1}(k) + \delta_{+1}(k) \right)}{2} \quad \Leftrightarrow \quad \hat{g}(\omega) = \cos \omega 
\end{equation}
\item Born and Jordan \cite{Born-Jordan_1925}
\begin{equation}\label{eq:QJP_Born_Jordan}
g(k) = 
\begin{cases}
\frac{1}{2}, \quad (|k| \leq 1) \\
0, \quad (|k| > 1)
\end{cases}
\quad \Leftrightarrow \quad 
\hat{g}(k) = \frac{\sin \omega}{\omega}
\end{equation}
\end{enumerate}
which all belongs to the same class generated by convolutions.

\paragraph*{The Theory of Hushimi and Glauber-Sudarshan}

In the previous paragraph, we have considered the problem of constructing various QJSDs of $\boldsymbol{X}$ by means of convolution.  In a broader perspective, however, the convolution $h \ast W_{\boldsymbol{X}}$ itself need not be a QJSD of $\boldsymbol{X}$ in the sense that it gives rise to the adjoint pair of representations in an extended sense.  If the map $h \ast: f \mapsto h \ast f$ is bijective, both the original QJSD and the convolution contain the same amount of information of $\boldsymbol{X}$, and a sufficient condition for this is $\hat{h} > 0$.

As for the choice of the function $h$, which may from the result \eqref{eq:conv_as_weighted_average} be interpreted as the `weight function' of the family of QJSDs, we typically consider the normal distribution ({\it i.e.}, Gau{\ss}ian function)
\begin{equation}
G(\boldsymbol{x}) := e^{- \|\boldsymbol{x}\|^{2}/2^{n}}, \quad \boldsymbol{x} \in \mathbb{R}^{n},
\end{equation}
where $\|\boldsymbol{x}\| := \sqrt{\sum_{i=1}^{n} |x_{i}|^{2}}$ denotes the Euclidean norm of an $n$-dimensional vector $\boldsymbol{x} \in \mathbb{R}^{n}$.  Since the normal distribution never satisfies the condition \eqref{eq:condition_conv_QJSD}, the convolution $G \ast \#_{\boldsymbol{X}}$ of a normal distribution with a QJSD of $\boldsymbol{X}$ is not a QJSD of $\boldsymbol{X}$ in general.  However, since the Fourier transform of a normal distribution is another normal distribution, hence $\hat{G} > 0$, the original QJSD $\#_{\boldsymbol{X}}$ and the convolution $G \ast \#_{\boldsymbol{X}}$ are equivalent to each other.  We thus introduce two operator valued distributions (OVDs) $H_{\boldsymbol{X}}$ and $GS_{\boldsymbol{X}}$ that are uniquely specified through the relations
\begin{align}
H_{\boldsymbol{X}} &= G \ast W_{\boldsymbol{X}}, \\
W_{\boldsymbol{X}} &= G \ast GS_{\boldsymbol{X}}.
\end{align}
By the above argument, we see that all $W_{\boldsymbol{X}}$, $H_{\boldsymbol{X}}$ and $GS_{\boldsymbol{X}}$ contain the same information in the sense that they can be transformed to one another by convolution with normal distributions, and thus the adjoint pairs of quantum/quasi-classical representations which they yield are all equivalent.  In this paper, we casually call the QJSD $W_{\boldsymbol{X}}$ the \emph{Wigner-Weyl} type, and the OVDs $H_{\boldsymbol{X}}$ and $GS_{\boldsymbol{X}}$ the \emph{Hushimi} and the \emph{Glauber-Sudarshan} type, respectively.  From the result \eqref{eq:convolution_transformation_classicalisation}, one sees that the quasi-classical representation of a density operator $\rho \in N(\mathcal{H})$ pertaining to the respective representations are related to each other by
\begin{align}
H_{\boldsymbol{X}}^{\rho} &= G \ast W_{\boldsymbol{X}}^{\rho}, \\
W_{\boldsymbol{X}}^{\rho} &= G \ast GS_{\boldsymbol{X}}^{\rho}.
\end{align}
As we have seen in the previous passage that $W_{\boldsymbol{X}}^{\rho}(q,p)$ is the Wigner function of $\rho$, we here learn that the distributions $H_{\boldsymbol{X}}^{\rho}(q,p)$ and $GS_{\boldsymbol{X}}^{\rho}(q,p)$ are precisely the \emph{Hushimi Q-function} \cite{Hushimi_1940} and the \emph{Glauber-Sudarshan P-function} \cite{Glauber_1963, Sudarshan_1963} of the quantum state $\rho$, respectively.

\newpage

\section{Some Applications: Quantum Correlations, Conditional Expectations and the Weak Value}\label{sec:three}

We now seek application of the formalisms we have constructed in the previous chapter.  By its nature, the framework of quantisation/quasi-classicalisation by QJSDs should become useful in analysing problems where non-commutativity of quantum observables is concerned.
In this section, we specifically focus on `correlations' and `conditioning' between (generally non-commuting) quantum observables induced by QJSDs.  Due to the non-commuting nature of quantum observables, there are various candidates of quantum correlations and conditional expectations.  Specifically, we see that Aharonov's weak value \cite{Aharonov_1964} can be identified as one realisation among various candidates of quantum conditional expectations.

\paragraph*{Complex Parametrised Subfamily}

In handling relatively abstract objects as QJSDs of quantum observables, concrete examples are always of use.  To this, we occasionally consider the simplest case where only two self-adjoint operators $\boldsymbol{A} = (A,B)$ are concerned, and make use of the complex parametrised subfamily of hashed operators
\begin{equation}\label{def:complex_param_QJSD}
\hat{\#}_{\boldsymbol{A}}^{\alpha}(s,t) := \frac{1 + \alpha}{2} \cdot e^{-itB}e^{-isA} + \frac{1 - \alpha}{2} \cdot e^{-isA}e^{-itB}, \quad  \alpha \in \mathbb{C},
\end{equation}
and the resulting subfamily $\#_{\boldsymbol{A}}^{\alpha} := \mathscr{F}^{-1}\hat{\#}_{\boldsymbol{A}}^{\alpha}$ of QJSDs for demonstration%
\footnote{
Do not confuse the subfamily introduced above with that of \eqref{def:QJSD_conv_subfamily}.  We have used different superscript characters $\kappa$, $\alpha$ as parameters to make the distinction more easier.}.

\subsection{Correlations of Quantum Observables}

In classical probability theory, \emph{correlation} has been an important quantity in various aspects.
The definition of the quantum counterpart, however, is not so obvious, when non-commutative observables are taken into account.
In what follows, we define a family of quantum correlation based on our framework of QJSDs of quantum observables, and observe their very basic properties.

\subsubsection{Sesquilinear Forms induced by QJP Distributions}

As usual, let $\boldsymbol{A} = (A_{1}, \dots, A_{n})$, $n \geq 1$, be an ordered combination of self-adjoint operators on a Hilbert space $\mathcal{H}$, and choose a QJSD $\#_{\boldsymbol{A}}$ and a density operator $\rho \in N(\mathcal{H})$.  In what follows, in order to refrain ourselves from dealing with unessential mathematical intricacies, we assume that the resulting QJP distribution $\rho_{\#_{\boldsymbol{A}}}$ can be represented by a density function%
\footnote{We say that a tempered distribution $u \in \mathscr{S}^{\prime}(\mathbb{K}^{n})$ admits representation by a density function if $u(x)$ is actually an integrable function.
}.
This allows us to introduce a sesquilinear form
\begin{equation}\label{def:quantum_correlation}
\langle g,\, f \rangle_{\rho_{\#_{\boldsymbol{A}}}} := \int_{\mathbb{K}^{n}} g^{\ast}(\boldsymbol{a}) f(\boldsymbol{a}) \ \rho_{\#_{\boldsymbol{A}}}(\boldsymbol{a}) dm_{n}(\boldsymbol{a}), \quad f,\ g \in L^{2}(\rho_{\#_{\boldsymbol{A}}})
\end{equation}
defined on the space of square-integrable functions with respect to the complex density function $\rho_{\#_{\boldsymbol{A}}}$.  By definition, we have
\begin{equation}
\langle f,\, g \rangle_{\rho_{\#_{\boldsymbol{A}}}} = \langle g^{\ast} ,\, f^{\ast} \rangle_{\rho_{\#_{\boldsymbol{A}}}}
\end{equation}
and
\begin{equation}
\langle f ,\, g \rangle_{\rho_{\#_{\boldsymbol{A}}}^{\ast}} = \langle g,\, f \rangle_{\rho_{\#_{\boldsymbol{A}}}}^{\ast},
\end{equation}
where the superscript asterisk denotes the complex conjugate.  The following observations are direct consequences of the above properties.
\begin{enumerate}
\item The quantum correlation \eqref{def:quantum_correlation} is \emph{symmetric} (Hermitian)
\begin{equation}
\langle f,\, g \rangle_{\rho_{\#_{\boldsymbol{A}}}} = \langle g,\, f \rangle_{\rho_{\#_{\boldsymbol{A}}}}^{\ast}
\end{equation}
if and only if the QJP distribution $\rho_{\#_{\boldsymbol{A}}}$ is \emph{real}.  This is guaranteed for every $\rho \in N(\mathcal{H})$ if and only if the choice of the QJPD $\#_{\boldsymbol{A}}$ is self-adjoint.
\item The quantum correlation \eqref{def:quantum_correlation} is \emph{positive definite}%
\footnote{
Here, the equality $f=0$ in the second line of \eqref{eq:positive-definiteness} is meant to hold $\rho_{\#_{\boldsymbol{A}}}$-almost everywhere.
}
\begin{equation}\label{eq:positive-definiteness}
\begin{split}
\forall f \in L^{2} \quad \langle f,\, f \rangle_{\rho_{\#_{\boldsymbol{A}}}} \geq 0, \\
\langle f,\, f \rangle_{\rho_{\#_{\boldsymbol{A}}}} = 0 \quad \Leftrightarrow \quad f = 0
\end{split}
\end{equation}
if and only if the QJP distribution $\rho_{\#_{\boldsymbol{A}}}$ is \emph{positive}.  This is guaranteed for every $\rho \in N(\mathcal{H})$ if and only if the QJPD $\#_{\boldsymbol{A}}$ is positive.
\end{enumerate}
Note that the second condition ({\it i.e.}, positive definiteness) is stronger that the first condition ({\it i.e.}, symmetricity), for indeed the positiveness of the QJP distributions trivially implies its realness, and in parallel, the positiveness of the QJPDs implies its self-adjointness.

\subsubsection{Quantum Correlations}

We next introduce the concept of quantum correlations based on the sesquilinear forms defined above.  In what follows, in order to ease our arguments, we confine ourselves to the simplest case $\boldsymbol{A} = (A,B)$ without loss of generality.  We also write $\# := \#_{\boldsymbol{A}}$ for better readability.  Now, let $f(A)$, $g(B)$ be operators respectively defined from the functions $f(a)$ and $g(b)$ by means of the functional calculus.  We then define the \emph{quantum correlations} or \emph{quasi-correlations} between the operators $f(A)$, $g(B)$ by
\begin{equation}
\langle g(B), f(A) \rangle_{\rho_{\#}} := \int_{\mathbb{K}^{2}} g^{*}(b)f(a)\ \rho_{\#}(a,b) dm_{2}(a,b)
\end{equation}
By construction, quantum correlations are dependent on the choice of the QJP distributions $\rho_{\#}$, which is generally non-unique due to the indefiniteness of the QJSDs.  When $A$ and $B$ happen to be simultaneously measurable, indefiniteness of the QJSDs vanishes, and the quantum correlation reduces to the unique classical correlation in the standard sense.

\subsubsection{Quantum Covariances}

Now that we have introduced the concept of quantum correlations, we next introduce the concept of quantum covariances.
Under the same assumptions, we introduce the \emph{quantum covariance} of the pair with respect to the QJP distribution $\rho_{\#}$ defined as
\begin{align}\label{def:quantum_covariance}
\mathbb{CV}[f(A),g(B);\rho_{\#}]
    &:= \langle g(B) - \mathbb{E}[g(B); \rho],\, f(A) - \mathbb{E}[f(A); \rho] \rangle_{\rho_{\#}} \nonumber \\
    &= \langle g(B),\, f(A) \rangle_{\rho_{\#}} - \mathbb{E}[f(A); \rho] \cdot \mathbb{E}[g(B); \rho],
\end{align}
where $\mathbb{E}[X; \rho] := \mathrm{Tr}[X\rho]$ denotes the expectation value of $X \in L(\mathcal{H})$ on a density operator $\rho \in N(\mathcal{H})$ as usual.
The quantum covariance serves as a natural extension to the standard covariance in classical probability theory, and they indeed coincide when the pair of self-adjoint operators $f(A)$ and $g(B)$ strongly commute.

\paragraph*{Example}

As an example, let $\#^{\alpha}$ be a complex parametrised QJSD for $\alpha \in \mathbb{C}$ introduced in \eqref{def:complex_param_QJSD}. By a simple computation, the quantum correlation of the operators $A$ and $B$ reads
\begin{align}
\langle B,\, A \rangle_{\rho_{\#}}
    &:= \int_{\mathbb{R}^{2}} ba\ \rho_{\#}(a,b)\ dm_{2}(a,b) \nonumber \\
    &= \left. (i\partial_{s_{1}})(i\partial_{s_{2}}) (\mathscr{F}\rho_{\#})(\boldsymbol{s}) \right|_{\boldsymbol{s} = 0} \nonumber \\
    &= \mathrm{Tr}\left[ \frac{\{A, B\}}{2} \rho \right] + i\,\alpha\, \mathrm{Tr}\left[ \frac{[A, B]}{2i} \rho \right],
\end{align}
where $\{X,Y\} := XY + YX$ and $[X,Y] := XY - YX$ respectively denotes the anti-commutator and the commutator of $X$ and $Y$ as usual.  This computation leads to
\begin{equation}\label{eq:quantum_covariance_example}
\mathbb{CV}[A, B;\rho_{\#^{\alpha}}]
    = \mathbb{CV}_{\mathrm{S}}[A, B;\rho] + i\,\alpha\, \mathbb{CV}_{\mathrm{A}}[A, B;\rho],
\end{equation}
where we have introduced the \emph{standard symmetric} and \emph{standard anti-symmetric} quantum covariances
\begin{align}
\mathbb{CV}_{\mathrm{S}}[A, B;\rho] &:= \mathrm{Tr}\left[ \frac{\{A, B\}}{2} \rho \right]  - \mathbb{E}[A; \rho] \cdot \mathbb{E}[B; \rho], \\
\mathbb{CV}_{\mathrm{A}}[A, B;\rho] &:= \mathrm{Tr}\left[ \frac{[A, B]}{2i} \rho \right],
\end{align}
for better readability.

\subsection{Conditioning by Quantum Observables}\label{sec:quantum_conditional_expectations}

In the previous passage, we have introduced quantum analogues of correlations by means of QJP distributions.  Closely related to these are quantum analogues of conditional expectations.

\subsubsection{Introducing Quantum Conditional Expectations}

In what follows, in order to avoid distraction by unessential mathematical intricacies, we impose an additional assumption that, for the given choice of the density operator $\rho \in N(\mathcal{H})$, the probability of finding the outcome of $B$ is always positive $\rho_{B}(b) > 0$ on its spectrum%
\footnote{
The spectrum of a self-adjoint operator $A$ is defined as the largest closed subset $J \subset \mathbb{R}$ such that  $E_{B}(J) = \mathrm{Id}$ holds.
}.
The quantum correlation of both the operators $f(A)$ and $g(B)$ then reads
\begin{align}\label{eq:QCE_Correlation}
\langle g(B),\ f(A) \rangle_{\rho_{\#}}
    &:= \int_{\mathbb{K}^{2}} g^{\ast}(b) f(a) \ \rho_{\#}(a,b) dm_{2}(a,b) \nonumber \\
    &= \int_{\mathbb{K}} g^{\ast}(b) \mathbb{E}[f(A) | B = b; \rho_{\#}] \ \rho_{B}(b) dm(b) \nonumber \\
    &= \langle g(B),\ \mathbb{E}[f(A) | B; \rho_{\#}] \rangle_{\rho_{B}},
\end{align}
where we have introduced the \emph{quantum conditional expectation} or \emph{conditional quasi-expectation}
\begin{equation}\label{def:quantum_conditional_expectation}
\mathbb{E}[f(A) | B = b; \rho_{\#}]
    := \frac{\int_{\mathbb{K}} f(a)\ \rho_{\#}(a,b) dm(a)}{\rho_{B}(b)}.
\end{equation}
of the operator $f(A)$ given the outcome of $B$ under the QJP distribution $\rho_{\#}$.
Note that the quantum conditional expectation $\mathbb{E}[f(A) | B; \rho_{\#}]$ is defined as an (equivalence class of) complex function(s) rather than a scalar.  The normal operator
\begin{equation}\label{def:quantum_conditional_expectation_operator}
\mathbb{E}[f(A) | B; \rho_{\#}] := \int_{\mathbb{K}} \mathbb{E}[f(A) | B = b; \rho_{\#}](b)\ dE_{B}(b)
\end{equation}
in the last equation is the image of the functional calculus of the (equivalence class of) function(s) \eqref{def:quantum_conditional_expectation}.

\paragraph*{Some `statistical' Properties and its Interpretation}

The key observation to make here is that the quantum correlation of an operator $f(A)$ with any operator $g(B)$ generated by $B$ can be reproduced by the \emph{authentic} correlation of the quantum conditional expectation $\mathbb{E}[f(B) | B; \rho_{\#}]$ with $g(B)$, which we reiterate for emphasis as:
\begin{equation}\label{eq:QCE_correlation_preserving}
\langle g(B),\ \mathbb{E}[f(A) | B; \rho_{\#}] \rangle_{\rho_{B}} = \langle g(B),\ f(A) \rangle_{\rho_{\#}}, \quad \forall g \in L^{2}(\rho_{B}).
\end{equation}
Also, by taking the constant function $g = 1$, we have
\begin{equation}
\mathbb{E}\left[ \mathbb{E}[f(A) | B; \rho_{\#}] ; \rho\right]  = \mathbb{E}\left[ f(A);\rho \right].
\end{equation}
The above two equalities show that the quantum conditional expectation serve as the `approximation' of the original operator $f(A)$ by operators generated by $B$, and that it is unique in the sense that it precisely reproduces the quantum correlation with any other operators generated by $B$ in place of the original $f(A)$.

In physical terms, the quantum conditional expectation can be interpreted as the quantum analogue of conditional expectations of the operator $f(A)$ given the outcome $b$ of $B$ under the hypothetical `joint' distribution ({\it i.e.}, QJP distribution) $\rho_{\#}$.  If the combination of the observables $\boldsymbol{A} = (A,B)$ happens to be simultaneously measurable, the quantum conditional expectation simply reduces to the conditional expectation in the classical sense that is familiar to us.  On the other hand, if some of the pair of observables fail to admit simultaneous measurability, the quantum conditional expectation becomes a hypothetical quantity, whose definition becomes non-unique due to the indefiniteness of the choice of the QJPDs.

\paragraph*{Examples}

We next provide some concrete examples to actually compute the quantum conditional expectations.  To this end, let $\#^{\alpha}$ be the parametrised subfamily of QJSDs of the ordered pair $\boldsymbol{A} = (A,B)$ of self-adjoint operators introduced in \eqref{def:complex_param_QJSD}.  
For a given function $f(a)$, let
\begin{equation}
f(a) = \sum_{m=0}^{\infty} f_{m}a^{m}
\end{equation}
denote its Taylor expansion.  If we let $\varphi(b)$ denote the numerator of \eqref{def:quantum_conditional_expectation}, we then have
\begin{align}
(\mathscr{F}\varphi)(t)
    &= \int_{\mathbb{R}} e^{-itb} \left( \int_{\mathbb{R}} \sum_{m=0}^{\infty} f_{m}a^{m}\ \rho_{\#^{\alpha}}(a,b) dm(a) \right) dm(b) \nonumber \\
    &= \sum_{m=0}^{\infty} f_{m} \left( \int_{\mathbb{R}^{2}} e^{-i(0a + tb)} a^{m} b^{0} \ \rho_{\#^{\alpha}}(a,b) dm_{2}(a,b) \right) \nonumber \\
    &= \sum_{m=0}^{\infty} f_{m} \left( (i\partial_{s})^{m}(i\partial_{t})^{0}\ (\mathscr{F}\rho_{\#^{\alpha}})(0,t) \right) \nonumber \\
    &= \sum_{m=0}^{\infty} f_{m} \left( \frac{1+\alpha}{2} \cdot  \mathrm{Tr}\left[ e^{-itB}A^{m} e^{-i0A} \rho \right] + \frac{1 -\alpha}{2} \cdot  \mathrm{Tr}\left[ A^{m} e^{-i0A}e^{-itB} \rho \right] \right) \nonumber \\
    &= \frac{1+\alpha}{2} \cdot \mathrm{Tr}\left[ e^{-itB} f(A) \rho \right] + \frac{1-\alpha}{2}  \cdot \mathrm{Tr}\left[ f(A)e^{-itB} \rho \right].
\end{align}
Observing that the Fourier transform of the function $b \mapsto \mathrm{Tr}\left[ E_{B}(b) f(A) \rho \right]$ and $b \mapsto \mathrm{Tr}\left[ f(A) E_{B}(b) \rho \right]$ are respectively $\mathrm{Tr}\left[ e^{-itB} f(A) \rho \right]$ and $\mathrm{Tr}\left[ f(A)e^{-itB} \rho \right]$, one finds that the injectivity and the linearity of the Fourier transformation leads to
\begin{equation}
\int_{\mathbb{R}} f(a)\ \rho_{\#^{\alpha}}(a,b) dm(a) = \frac{1+\alpha}{2} \cdot \mathrm{Tr}\left[ E_{B}(b) f(A) \rho \right] + \frac{1-\alpha}{2}  \cdot \mathrm{Tr}\left[ f(A) E_{B}(b) \rho \right]
\end{equation}
by combining the results.  We thus finally have
\begin{align}\label{eq:QCE_example}
\mathbb{E}^{\alpha}[A | B=b ; \rho]
    &:= \mathbb{E}[f(A) | B=b; \rho_{\#^{\alpha}}] \nonumber \\
    &= \frac{1+\alpha}{2} \cdot \frac{\mathrm{Tr}\left[ E_{B}(b) f(A) \rho \right]}{\mathrm{Tr}[E_{B}(b)\rho]} + \frac{1-\alpha}{2}  \cdot \frac{\mathrm{Tr}\left[ f(A) E_{B}(b) \rho \right]}{\mathrm{Tr}[E_{B}(b)\rho]} \nonumber \\
    &= \mathrm{Re}\,\left[ \frac{\mathrm{Tr}\left[ E_{B}(b) f(A) \rho \right]}{\mathrm{Tr}[E_{B}(b)\rho]} \right] + i\,\alpha\, \mathrm{Im}\,\left[ \frac{\mathrm{Tr}\left[ E_{B}(b) f(A) \rho \right]}{\mathrm{Tr}[E_{B}(b)\rho]} \right],
\end{align}
where we have introduced an abbreviated symbol (the first line) for better readability.

\subsection{The Weak Value as a Quantum Conditional Expectation}

As the simplest case of the examples provided in the previous Section~\ref{sec:quantum_conditional_expectations}, let us consider the quantum conditional expectation of the function $f(a,b) = a$.  Based on the formula \eqref{eq:QCE_example}, one obtains
\begin{align}\label{def:Two_state_value}
\mathbb{E}^{\alpha}[A | B=b ; \rho]
    &= \frac{1+\alpha}{2} \cdot A_{w}^{\rho}(b) + \frac{1-\alpha}{2}  \cdot A_{w}^{\rho}(b)^{*} \nonumber \\
    &= \mathrm{Re}\,\left[ A_{w}^{\rho}(b) \right] + i\,\alpha\, \mathrm{Im}\,\left[ A_{w}^{\rho}(b) \right], \quad \alpha \in \mathbb{C},
\end{align}
where we have introduced the Aharonov's weak value
\begin{equation}\label{def:Weak_Value}
A_{w}^{\rho}(b)
    := \mathbb{E}^{1}[A | B = b; \rho] \nonumber \\
    = \frac{\mathrm{Tr}\left[ E_{B}(b) A \rho \right]}{\mathrm{Tr}[E_{B}(b)\rho]}.
\end{equation}
The quantity \eqref{def:Two_state_value}, defined as the complex convex combination of the weak value, was initially proposed in \cite{Morita_2013}, in which it was called the \emph{two state value}.  Provided that the density operator $\rho = |\psi\rangle\langle\psi|$ is an orthogonal projection onto the $1$-dimensional linear subspace spanned by a unit vector $|\psi\rangle \in \mathcal{H}$, weak value reduces to its familiar form
\begin{equation}
A_{w}^{\psi}(b) = \frac{\langle b | A | \psi \rangle}{\langle b | \psi \rangle}.
\end{equation}
In this respect, the weak value admits an interpretation as one manifestation of the possible family of quantum conditional expectations corresponding to the specific choice $\#^{\alpha}$, $\alpha = 1$, of the subfamily of the QJSDs.

The interpretation of the weak value, as one of the possible candidates of quantum analogues of classical conditional expectations, has been proposed earlier in several literatures.  Relevant to our framework presented here is \cite{Dressel_2015_03}, in which the inherent non-uniqueness of the QJP distributions for non-commuting observables is particularly emphasised, a fact which is oftentimes overlooked in discussing the weak value. There, quantum conditional expectations are computed not only for the Kirkwood-Dirac distribution \eqref{eq:QJP_Kirkwood_Dirac}, but also for several other types of QJP distributions, including the Wigner-Weyl distribution \eqref{eq:QJP_Weyl_Wigner} and the Margenau-Hill distribution \eqref{eq:QJP_Margenau_Hill}.

\section{Summary and Discussion}

In the former part of this paper (Section~\ref{sec:two}), we focused on the problem of \emph{quantisation} of classical observables and \emph{quasi-classicalisation} of quantum states.  For the simplest case in which the observables concerned are all simultaneously measurable, we reviewed that the joint-spectral measure (JSM) uniquely attributed to the commuting observables gives rise to the unique pair of functional analysis and the Born rule, which could respectively be considered as the trivial realisation of quantisation and quasi-classicalisation.  Specifically, by taking the duality relation between observables and states into account, we saw that quantisation and quasi-classicalisation are actually adjoint to one another as maps, and thus the JSM, quantisation and quasi-classicalisation are all equivalent as an entity, although they may differ in concept.  In this sense, we occasionally referred to the pair of maps as quantum/quasi-classical transformations or representations.

We next considered the general case in which observables concerned are arbitrary.  To this, we let ourselves be guided by the observation above, and introduced the concept of \emph{quasi-joint-spectral distributions} (QJSDs), which could be interpreted as non-commutative analogues to the standard JSM.  In contrast to the commutative case, QJSDs attributed to a given set of non-commuting observables are non-unique, and thus they give rise to various distinct pairs of quantisations and quasi-classicalisations.  We also discussed the basic properties of QJSDs and the transformations between them.  An important implication of this framework is that, although there may be countless possible ways to construct quantisation and quasi-classicalisation, there is a precise one-to-one correspondence between them.  These realisations help us to understand the relation between various proposals made historically,  
including those proposed by Wigner-Weyl, Kirkwood-Dirac, Margenau-Hill, Born-Jordan, Hushimi and Glauber-Sudarshan.

As an application to this framework, the latter portion of this paper (Section~\ref{sec:three}) focused on the problem of constructing quantum analogues to the classical concept of correlation and conditioning.  We proposed a framework to this problem by means of QJSDs introduced earlier, and demonstrated that some of the statistical properties familiar in classical probability theory are still preserved even under the quantum counterpart, especially the relation between correlation and conditional expectation.  We finally mentioned that Aharonov's weak value could be interpreted as one manifestation of quantum conditional expectations.  One of the virtues of this interpretation is that it reveals a novel aspect of the uncertainty relations \cite{Lee_2016}, in the sense that the weak value appears as the optimal choice of approximation: quantum conditional expectations are best approximations of an observable by another observable, just as classical conditional expectation is the best approximation of a random variable by means of another.

The framework of the quantum/quasi-classical transformation proposed in this paper may find a variety of applications.  In fact, it should be obvious from our arguments that one can always draw an analogy to various concepts and results in classical probability theory when one considers the quantum counterparts obtained by this method.  Naturally, this will allow for an intuitive treatment of the latter based on the statistical and geometric structures present in classical probability theory.

\begin{acknowledgements}
This work was supported by MEXT Grant-in-Aid for Scientific Research on Innovative Areas `Science of hybrid quantum systems' (Grant No. 15H05870).
\end{acknowledgements}

\bibliographystyle{unsrt}
\bibliography{main}

\end{document}